\documentclass[reprint,showpacs,aps,amsmath,amssymb,prd,nofootinbib,floatfix,superscriptaddress]{revtex4-1} 

\usepackage{graphicx}
\usepackage{bm}
\usepackage{dcolumn}
\usepackage{ulem} 
\usepackage{float}
\usepackage[usenames]{color}
\usepackage{latexsym}

\usepackage{times}
\usepackage{mathptmx}

\usepackage{amssymb,amsmath}

\begin{document}



\title{Quasiequilibrium sequences of binary neutron stars undergoing dynamical scalarization}


\author{Keisuke Taniguchi}
\affiliation{Graduate School of Arts and Sciences,
University of Tokyo, Komaba, Meguro, Tokyo, 153-8902, Japan}


\author{Masaru Shibata}
\affiliation{Yukawa Institute for Theoretical Physics,
Kyoto University, Kyoto, 606-8502, Japan}

\author{Alessandra Buonanno}
\affiliation{Max Planck Institute for Gravitational Physics
(Albert Einstein Institute), 
Am M\"uhlenberg 1, Potsdam-Golm, 14476, Germany}
\affiliation{Department of Physics, University of Maryland, College Park,
Maryland 20742, USA}

\date{\today}

\begin{abstract}
We calculate quasiequilibrium sequences of equal-mass,
irrotational binary neutron stars in a scalar-tensor theory
of gravity that admits dynamical scalarization.
We model neutron stars with realistic equations of state
(notably through piecewise polytropic equations of state).
Using these quasiequilibrium sequences we compute the binary's scalar charge
and binding energy versus orbital angular frequency.
We find that the absolute value of the binding energy
is smaller than in general relativity, differing at most by $\sim 14 \%$ 
at high frequencies for the cases considered. 
We use the newly computed binding energy and the balance equation
to estimate the number of gravitational-wave (GW) cycles during the adiabatic,
quasicircular inspiral stage up to the end of the sequence, which is 
the last stable orbit or the mass-shedding point, depending on which
comes first. We find that, depending on the scalar-tensor parameters,
the number of GW cycles can be substantially smaller than in general relativity.
In particular, we obtain that when dynamical scalarization sets in around 
a GW frequency of $\sim 130 \,{\rm Hz}$, the sole inclusion of the
scalar-tensor binding energy causes a reduction of GW cycles from
$\sim 120 \,{\rm Hz}$ up to the end of the sequence ($\sim 1200\,{\rm Hz}$) of 
$\sim 11 \%$ with respect to the general-relativity case.
(The number of GW cycles from $\sim 120 \,{\rm Hz}$ to the end of the sequence 
in general relativity is $\sim 270$.) We estimate that when the scalar-tensor 
energy flux is also included the reduction in GW cycles becomes of
$\sim 24 \%$. 
Quite interestingly, dynamical scalarization can produce a difference
in the number of GW cycles with respect to the general-relativity
point-particle case that is much larger than the effect due to tidal
interactions, which is on the order of only a few GW cycles.
These results further clarify and confirm recent studies that 
have evolved binary neutron stars either in full numerical relativity
or in post-Newtonian theory,
and point out the importance of developing accurate scalar-tensor-theory
waveforms for systems composed of strongly self-gravitating objects,
such as binary neutron stars.
\end{abstract}


\maketitle

\section{Introduction}
\label{intro}

Coalescing binary neutron stars are among the most promising sources
of gravitational waves (GWs) for the next-generation, kilometer-size, 
GW detectors such as advanced LIGO~\cite{LIGO10},
advanced Virgo~\cite{VIRGO11}, and KAGRA~\cite{KAGRA10}.
Binary neutron stars, together with black-hole$-$neutron-star binaries,
are also regarded as one of the candidate central engines of
short-hard gamma-ray bursts~\cite{NarayPP92}.
The use of matched-filtering technique to detect GW 
signals from coalescing binary systems and the interest in shedding
light on gamma-ray burst progenitors have led to impressive advances
in modeling the dynamics and gravitational waveforms of binary
neutron stars (see, e.g., Ref.~\cite{Blanchet06} for the inspiral phase and
Refs.~\cite{HotokKOSK11,HotokKKMSST13,BernuDWB13} for the merger
and postmerger phases).
Most of those studies were carried out in general relativity,
except for Refs.~\cite{BarauPPL13,ShibaTOB14,PalenBPL14}.
Although general relativity has passed all known experimental
and observational tests in the weak-field and slow-motion
limit (see e.g., Ref.~\cite{Will05}), it remains to be seen whether it will 
survive tests in the strong-field and high-velocity regime.

The detection of GWs emitted by coalescing binary
systems offers the unique opportunity to investigate the validity
of general relativity in the strong-field regime.
To achieve this goal, accurate gravitational waveforms in gravity theories
alternative to general relativity~\cite{Will93,Will05,WillZ89} need
to be computed.
Here we follow our recent work~\cite{ShibaTOB14} and
focus on the scalar-tensor model~\cite{Jordan49,Fierz56,BransD61,FujiiM03}
proposed by Damour and Esposito-Far\`ese (DEF)~\cite{DamourEF93}
(see also Refs.~\cite{DamourEF92,DamourEF96b,DamourEF98}).
Quite interestingly, there exist choices of the free parameters
in the DEF model, for which both weak and mildly strong gravitational
tests are satisfied, notably the pulsar timing tests,
but strong-field tests could be violated and these violations could be
observed through the emission of GWs from the last stages of the binary's
inspiral, plunge, and merger in advanced LIGO, Virgo, and KAGRA detectors.
This is possible because if neutron stars in binary systems carry
negligible scalar charge when largely separated,
they can be dynamically scalarized as they come closer to each other
through gravitational interaction, i.e., they undergo dynamical scalarization
as the binary's compactness increases~\cite{BarauPPL13,ShibaTOB14,PalenBPL14}.
[See also Refs.~\cite{DonevaYSK13,PaniB14,DonevaYSKA14} for scalarization
of rotating stars.] 
However, it is important to notice that for the same values of the DEF
parameters for which dynamical scalarization can occur, the DEF model 
may have problems in providing cosmological solutions consistent with
our Universe~\cite{DamourN93a,DamourN93b,Sampetal14}. It will be relevant 
to further investigate this problem in the future.

Barausse {\it et al.}~\cite{BarauPPL13} showed that dynamical scalarization
takes place in the DEF model by performing numerical-relativity 
simulations of inspiraling binary neutron stars. They performed two 
numerical simulations, which differed by the strength of the scalar field  
and the binary's mass ratio. Their simulations used approximate initial data,  
i.e., initial data computed by numerical codes of general relativity instead of 
the ones of scalar-tensor theory, and employed the polytropic 
equation of state $p/c^2 =K \rho_0^{\Gamma}$ with $\Gamma=2$ and
$K=123 G^3 M_{\odot}^2/c^6$, where $p$ is the pressure and $\rho_0$
is the baryonic rest-mass density in their notation.
For the unequal-mass binary the individual baryonic rest masses were
$1.78 M_{\odot}$ and $1.90 M_{\odot}$,
while for  the equal-mass binary the baryonic rest mass 
was $1.625 M_{\odot}$.\footnote{We notice that due to a misleading output
in the \textsc{LORENE} data set~\cite{lorene}, the gravitational masses  
are not the ones reported in Ref.~\cite{BarauPPL13}.
The gravitational masses of spherical, isolated stars corresponding
to the baryonic rest masses of $1.625 M_{\odot}$, $1.78 M_{\odot}$, and
$1.90 M_{\odot}$ are $1.51 M_{\odot}$, $1.64 M_{\odot}$, and $1.74 M_{\odot}$,
respectively. The results of Ref.~\cite{BarauPPL13} are correct; i.e.,
they were not affected by the gravitational masses reported in the paper.}

For comparison, we computed not in general relativity but in the DEF
scalar-tensor theory the initial data for the same baryonic rest masses
used in Barausse {\it et al.}. 
We set $\beta/(4 \pi G)=-4.5$ and $\varphi_{0,\rm{BPPL}} = 10^{-5} G^{-1/2}$,
where $\beta$ is a constant related to the derivative of a
scalar field and $\varphi_{0,\rm{BPPL}}$ is the asymptotic value of
the scalar field as defined by Barausse {\it et al.}~\cite{BarauPPL13}.
We found that the more massive star (with baryonic rest mass of $1.90 M_{\odot}$)
is spontaneously scalarized for a spherical configuration, and thus,
the unequal-mass binary system is already scalarized at the orbital separation
of $60$ km, which is where Ref.~\cite{BarauPPL13} starts their simulations
(see Appendix~\ref{app1} for more details).
Because in their simulation the scalar field does not exist initially, 
it rapidly increases just after the simulation starts.
This artificial behavior may have left imprints in the dynamical evolution
of the binary system. Indeed, as we will see in Sec.~\ref{sec3}, the binding 
energy computed along a sequence of quasiequilibrium configurations
in the DEF scalar-tensor theory is in absolute value smaller than
in general relativity. Thus, as soon as the simulation starts,
if initial data without the scalar field are used, which is the case
in Barausse {\it et al.}, then the absolute value of the binding 
energy will become smaller because the scalar field increases.
This means that the initial datum is a local minimum of the binding energy
along the quasiequilibrium sequences; i.e., it is a local turning 
point of the binding energy. We suspect that the fast plunge seen
in Ref.~\cite{BarauPPL13} might be enhanced by this effect. Performing 
a numerical simulation using general-relativity and scalar-tensor-theory 
initial data will clarify this point.

For the equal-mass binary system numerically evolved in Ref.~\cite{BarauPPL13}
we found, using quasiequilibrium configurations, that the binary is
already dynamically scalarized at the orbital separation
of $40$ km where Ref.~\cite{BarauPPL13} observed dynamical scalarization.
However, our result of an earlier dynamical scalarization may not be
in contradiction with Ref.~\cite{BarauPPL13}. 
Indeed, typically we found that the onset of dynamical scalarization
in the quasiequilibrium-configuration study occurs earlier than that
in dynamical simulation~\cite{ShibaTOB14}. This discrepancy may occur 
due to the breakdown of the assumption of quasiequilibrium. For a few 
orbits before merger the infall velocity of each star in the binary system is 
larger in numerical-relativity simulations than in quasiequilibrium 
configurations. As a consequence, in numerical-relativity simulations 
the binary can merge before the scalar field reaches its equilibrium state,
while, in the quasiequilibrium situation, the scalar field can reach
its equilibrium state even just before the quasiequilibrium sequence ends.
Thus, the effect of the scalar field is overestimated in the quasiequilibrium
study for the cases in which dynamical scalarization occurs just before the
end of the quasiequilibrium sequences. We will present more details about 
the results of the equal-mass binary in Appendix~\ref{app1}.

More recently, Palenzuela {\it et al.}~\cite{PalenBPL14} investigated
analytically the phenomenon of dynamical scalarization in the DEF model.
They employed the 2.5 post-Newtonian (PN) equations of motion, recently
derived in Ref.~\cite{MirshW13}, augmented by a set of equations that
phenomenologically describe the increase of scalar charge as the two
neutron stars come closer to each other. In this analysis the binary neutron
stars are approximated by two isolated, spherical neutron stars.
Reference~\cite{PalenBPL14} confirmed and quantified what was found
in Refs.~\cite{BarauPPL13,ShibaTOB14}, notably the fact that binary neutron
stars plunge and merge in the DEF model sooner than in general relativity
when they undergo induced and dynamical scalarization.

To further understand the onset of dynamical scalarization 
during the last stages of inspiral, we disentangle conservative from
radiative effects and compute for the first time quasiequilibrium sequences
of binary neutron stars in the DEF model. Our motivations are threefold.
We want to (i) produce initial data for merger simulations in the
scalar-tensor model~\cite{ShibaTOB14},
(ii) accurately extract physical quantities (notably the binding energy
and angular momentum) during the last stages of inspiral, just before merger,
where the effect of gravity becomes strong and the finite-size effect of
a neutron star starts to affect the evolution of the binary system,
and (iii) use those quantities to estimate by how much
the gravitational waveforms in the DEF model 
differ from the ones in general relativity.

This paper is organized as follows.
In Sec.~II, we give a brief summary of the quasiequilibrium-sequence
formalism. The formulation is based on the conformal thin-sandwich
decomposition. In Sec.~III we present and discuss the numerical results of
the scalar charge and scalar mass, binding energy, total angular momentum, 
central baryonic rest-mass density, evolution of the orbital angular frequency,
and number of GW cycles. Section IV summarizes
our main conclusions.

Throughout this paper, we employ the geometrical units of
$c=G=1$, where $c$ is the speed of light and $G$ is the bare gravitational 
constant. We use greek letters to denote spacetime components and latin
letters for the spatial components.

\section{Formulation} \label{sec:form}

As in Ref.~\cite{ShibaTOB14}, we work in the Jordan
frame~\cite{Jordan49,BransD61}.
The basic field equations for computing the metric quantities and
the scalar field are derived by taking variation of the action,
\begin{eqnarray}
  {\cal S} &=&\frac{1}{16\pi}
  \int \Bigl[ \phi R -\omega (\phi) \phi^{-1} g^{\mu \nu}
    (\nabla_{\mu} \phi) (\nabla_{\nu} \phi) \Bigr] \sqrt{-g} d^4 x
  \nonumber \\
  &&+{\cal S}_{\rm matter},
\label{action}
\end{eqnarray}
where $\phi$ is the scalar field, $g_{\mu \nu}$ is the spacetime metric
in the Jordan frame, $R$ is the Ricci scalar calculated
from $g_{\mu \nu}$, $g$ is the determinant of $g_{\mu \nu}$,
$\nabla_{\mu}$ is the covariant derivative with respect to $g_{\mu \nu}$,
and ${\cal S}_{\rm matter}$ is the matter part of the action.
The quantity $\omega$ is a function of $\phi$ that takes the form 
\begin{eqnarray}
  \frac{1}{\omega(\phi) +3/2} =B \ln \phi
\end{eqnarray}
in the DEF theory, where $B$ is a free parameter~\cite{DamourEF93}
(see Ref.~\cite{ShibaTOB14} for more details). 
The relation between Newton's constant $G_{\rm N}$ and
the bare gravitational constant $G$ is 
\begin{equation}
  G_{\rm N} = \frac{G}{\phi_0} \frac{4 +2 \omega(\phi_0)}{3 +2\omega(\phi_0)},
\end{equation}
where $\phi_0$ is the value of $\phi$ at spatial infinity.
For the values used in this paper, the deviation of the ratio $G_{\rm N}/G$
from unity is on the order of $10^{-9} - 10^{-10}$
[see Eq.~(\ref{eq:varphi_phi}) and the scalar-tensor 
values listed at the end of Sec.~\ref{sec:form}].

Taking variation of the action (\ref{action}) with respect to the metric
and the scalar field, we obtain
\begin{eqnarray}
  R_{\mu \nu} -\frac{1}{2} R g_{\mu \nu} &=&8 \pi \phi^{-1} T_{\mu \nu}
  +\omega \phi^{-2} \Bigl[ (\nabla_{\mu} \phi) (\nabla_{\mu} \phi) \Bigr.
    \nonumber \\
    &&\Bigl. -\frac{1}{2} g_{\mu \nu} (\nabla_{\alpha} \phi)
    (\nabla^{\alpha} \phi) \Bigr] \nonumber \\
  &&+\phi^{-1} \bigl( \nabla_{\mu} \nabla_{\nu} \phi -g_{\mu \nu} \Box \phi \bigr)
  \label{eq:metric}
\end{eqnarray}
and 
\begin{equation}
  \Box \phi =\frac{1}{2 \omega +3} \Bigl[ 8\pi T -\frac{d \omega}{d \phi}
    (\nabla_{\mu} \phi) (\nabla^{\mu} \phi) \Bigr], \label{eq:scalar}
\end{equation}
respectively, where $R_{\mu \nu}$ is the Ricci tensor, $\Box$ is
$\nabla_{\mu} \nabla^{\mu}$, and $T_{\mu \nu}$ is the stress-energy tensor.
For an ideal fluid we have 
\begin{equation}
  T_{\mu \nu} = (\rho +\rho \epsilon +P) u_{\mu} u_{\nu} +P g_{\mu \nu},
\end{equation}
where $u_{\mu}$ is the fluid 4-velocity, $\rho$ is the baryonic rest-mass
density, $\epsilon$ is the specific internal energy,
and $P$ is the pressure. Then, we set the metric line element in 3+1 form,
\begin{eqnarray}
  ds^2 &=& g_{\mu \nu} dx^{\mu} dx^{\nu}, \nonumber \\
  &=& -\alpha^2 dt^2 +\gamma_{ij} (dx^i +\beta^i dt)(dx^j +\beta^j dt)
\end{eqnarray}
where $\alpha$ is the lapse function, $\beta^i$ is the shift vector,
and $\gamma_{ij}$ is the spatial part of the spacetime metric, and 
we solve the basic field equations in the conformal thin-sandwich
decomposition~\cite{York99,PfeifY03}. We decompose the equations for 
the metric quantities (\ref{eq:metric}) into the {\it Hamiltonian} constraint,
\begin{eqnarray}
  {}^{(3)}R +K^2 -K_{ij} k^{ij} &=&16 \pi \phi^{-1} \rho_{\rm h} \nonumber \\
  &&+\omega \phi^{-2} \bigl[ \Pi^2 +(D_{\alpha} \phi)(D^{\alpha} \phi) \bigr]
  \nonumber \\
  &&+2 \phi^{-1} (D_{\mu} D^{\mu} \phi -K \Pi),
\end{eqnarray}
and the {\it momentum} constraint,
\begin{eqnarray}
  D_i K^i_j -D_j K &=&8 \pi \phi^{-1} J_j +\omega \phi^{-2} \Pi D_j \phi
  \nonumber \\
  &&+\phi^{-1} (D_j \Pi -K_j^i D_i \phi);
\end{eqnarray}
furthermore the trace part of the evolution equation for the extrinsic
curvature $K_{ij}$ satisfies the following equation
\begin{eqnarray}
  &&(\partial_t -\beta^k \partial_k) K =
  4 \pi \alpha \phi^{-1} (\rho_{\rm h} +S) +\alpha K_{ij} K^{ij}
  -D_i D^i \alpha \nonumber \\
  &&\hspace{25pt}+\alpha \omega \phi^{-2} \Pi^2 +\alpha \phi^{-1} D_i D^i \phi
  -\alpha \phi^{-1} K \Pi \nonumber \\    
  &&\hspace{25pt}-\frac{3\alpha \phi^{-1}}{2(2\omega +3)} \Bigl[ 8\pi T
    +\frac{d\omega}{d\phi} \Bigl\{ \Pi^2 -(D_k \phi) (D^k \phi) \Bigr\} \Bigr],
  \nonumber \\
\end{eqnarray}
while the evolution equation for the spatial metric reads
\begin{eqnarray}
  \partial_t \gamma_{ij} =-2\alpha K_{ij} +\gamma_{kj} D_i \beta^k
  +\gamma_{ik} D_j \beta^k,
\end{eqnarray}
where ${}^{(3)}R$ denotes the Ricci scalar calculated from $\gamma_{ij}$,
$D_i$ the covariant derivative with respect to $\gamma_{ij}$,
$K$ the trace part of the extrinsic curvature, and
$\Pi \equiv -n^{\mu} \partial_{\mu} \phi$.
Here the quantities $\rho_{\rm h}$, $J_i$, $S$, and $T$ are defined as
\begin{subequations}
\begin{eqnarray}
  \rho_{\rm h} &=& n_{\mu} n_{\nu} T^{\mu \nu}, \\
  J_i &=& -n_{\mu} \gamma_{\nu i} T^{\mu \nu}, \\
  S &=& \gamma^{ij} (\gamma_{i \mu} \gamma_{j \nu} T^{\mu \nu}), \\
  T &=& g_{\mu \nu} T^{\mu \nu},
\end{eqnarray}
\end{subequations}
where $n^{\mu}$ is the unit normal to the spatial hypersurface.

In the above decomposition, there appear four freely specified quantities:
the background spatial metric, $\tilde{\gamma}_{ij}$,
the time derivative of the background spatial metric in
contravariant form, $\partial_t \tilde{\gamma}^{ij}$,
the trace part of the extrinsic curvature, $K$, and
its time derivative, $\partial_t K$.
The background spatial metric is defined by
$\tilde{\gamma}_{ij} \equiv \psi^{-4} \gamma_{ij}$
where $\psi$ is the conformal factor.
Since we consider a stationary state, we set to zero the time derivatives
of the above quantities.
We also set to zero the trace part of the extrinsic curvature, $K$,
because we impose the condition of maximal slicing.
We further require that the background spatial metric, $\tilde{\gamma}_{ij}$,
be flat; that is, $\tilde{\gamma}_{ij} =f_{ij}$ where
$f_{ij}$ is the flat spatial metric~\cite{Isenb78,Isenb08,WilsoM89}.

The equation for the scalar field (\ref{eq:scalar}) is also rewritten
in the conformal thin-sandwich decomposition as
\begin{eqnarray}
  &&(\partial_t -\beta^i \partial_i) \Pi =-\alpha D_i D^i \phi
  -(D_i \alpha) (D^i \phi) +\alpha K \Pi \nonumber \\
  &&\hspace{25pt}+\frac{\alpha}{2 \omega +3}
  \Bigl[ 8 \pi T +\frac{d\omega}{d\phi} \bigl( \Pi^2
    -(D_k \phi) (D^k \phi) \bigr) \Bigr].
\end{eqnarray}
The above equation depends on the quantities $\Pi$ and $\partial_t \Pi$. 
Since we consider a stationary state, we set to zero $\partial_t \Pi$. 
For the quantity $\Pi$, we need to guarantee that it behaves at least
as $\Pi ={\cal O} (r^{-2})$ in the far zone. 
This is because the right-hand side of the Hamiltonian constraint and
the trace part of the evolution equation for $K_{ij}$
should decrease fast enough to ensure the spacetime to be
asymptotically flat. In this paper, for simplicity, we set to zero the 
quantity $\Pi$ (see Sec.~II.D of Ref.~\cite{ShibaTOB14} for a more detailed 
discussion).

Note that, as we mentioned above, we have the freedom of choosing another
background spatial metric, $\tilde{\gamma}_{ij}$, as well as
the quantity $\Pi$, and the trace part of the extrinsic curvature, $K$.
We think that the choice we made for the background spatial metric 
does not affect the main results of this paper, notably the onset of 
dynamical scalarization along quasiequilibrium binary neutron stars. 
As we will see in Sec.~\ref{sec:scalar}, the location of dynamical 
scalarization, i.e., the orbital angular frequency at the onset of
dynamical scalarization, agrees with what determined by fully relativistic
simulations and estimated by the analytical method discussed in
Ref.~\cite{ShibaTOB14}.
Because the simulations and the analytical estimation do not rely 
on the assumptions used in this paper, our results for the 
location of dynamical scalarization can be considered robust.

Thus, the equations for the quantities that enter in the metric can be
recast in the form
\begin{widetext}
\begin{subequations}
\begin{eqnarray}
  \Delta \psi &=& -2\pi \exp( -\varphi^2/2 ) \psi^5 \rho_{\rm h}
  -\frac{1}{8} \psi^{-7} \tilde{A}_{ij} \tilde{A}^{ij}
  -\frac{1}{2} \pi B \psi^5 \varphi^2 T \exp( -\varphi^2/2 )
  -\frac{1}{4} \psi \Bigl( 1 +\frac{1}{B} -\frac{3}{4} \varphi^2 \Bigr)
  f^{ij} (\partial_i \varphi) (\partial_j \varphi) \nonumber \\
  && +\frac{1}{4} \Phi^{-1} \varphi f^{ij} (\psi \partial_i \Phi
  -\Phi \partial_i \psi) (\partial_j \varphi), \label{eq:deltapsi} \\
  \Delta \Phi &=& 2 \pi \exp( -\varphi^2/2 ) \Phi \psi^4 (\rho_{\rm h} +2S)
  +\frac{7}{8} \Phi \psi^{-8} \tilde{A}_{ij} \tilde{A}^{ij}
  -\frac{3}{2} \pi B \Phi \psi^4 \varphi^2 T \exp( -\varphi^2/2 )
  -\frac{1}{4} \Phi \Bigl( 3 +\frac{1}{B} -\frac{3}{4} \varphi^2 \Bigr)
  f^{ij} (\partial_i \varphi) (\partial_j \varphi) \nonumber \\
  && -\frac{3}{4} \psi^{-1} \varphi f^{ij} (\psi \partial_i \Phi
  -\Phi \partial_i \psi) (\partial_j \varphi), \\
  \Delta \beta^i &+&\frac{1}{3} f^{ij} \partial_j (\partial_k \beta^k)
  = 16 \pi \exp( -\varphi^2/2 ) \Phi \psi^{-1} f^{ij} J_j -2\Phi \psi^{-7} \tilde{A}^{ij}
  (7 \psi^{-1} \partial_j \psi -\Phi^{-1} \partial_j \Phi)
  -2 \Phi \varphi \psi^{-7} \tilde{A}^{ij} \partial_j \varphi, \\
  \tilde{A}^{ij} &=&\frac{\psi^7}{2\Phi} \Bigl( f^{kj} \partial_k \beta^i
  +f^{ik} \partial_k \beta^j
  +\frac{2}{3} f^{ij} \partial_k \beta^k \Bigr),
\end{eqnarray}
\end{subequations}
\end{widetext}
where $\tilde{A}^{ij}$ is the traceless part of the conformal
extrinsic curvature defined as
\begin{equation}
  \tilde{A}^{ij} =\psi^{10} \Bigl(K^{ij} -\frac{1}{3} \gamma^{ij} K \Bigr),
\end{equation}
the quantity $\Phi$ is defined as $\Phi \equiv \alpha \psi$, and
we introduce a new scalar field $\varphi$ which is related to the 
scalar field $\phi$ as
\begin{equation}
  \varphi \equiv \sqrt{2 \ln \phi}. \label{eq:varphi_phi}
\end{equation}
The equation for the scalar field (\ref{eq:scalar}) can be rewritten
in the conformal thin-sandwich decomposition imposing
$K=0$, $\partial_t \Pi =0$, and $\Pi=0$, as
\begin{eqnarray}
  \Delta \varphi &=& 2\pi B \psi^4 \varphi T \exp( -\varphi^2/2 )
  -\varphi f^{ij} (\partial_i \varphi) (\partial_j \varphi) \nonumber \\
  &&-f^{ij} (\Phi^{-1} \partial_i \Phi +\psi^{-1} \partial_i \psi)
  (\partial_j \varphi).
\end{eqnarray}
(See Ref.~\cite{ShibaTOB14} for a more detailed derivation of the equations
for the quantities entering the metric and the scalar field.)

We also need to solve for the relativistic hydrostatic equations,
$\nabla_{\mu} T^{\mu \nu}=0$, which are basically the same as
in general relativity because they are written in the Jordan frame.
The equations are decomposed into the first integral of the Euler equation,
\begin{equation}
  h \alpha \frac{\gamma}{\gamma_0} = \textrm{const},
\end{equation}
and the equation of continuity,
\begin{equation}
  \frac{\rho}{h} \nabla_{\mu} \nabla^{\mu} \Psi
  +(\nabla_{\mu} \Psi) \nabla^{\mu} \Bigl( \frac{\rho}{h} \Bigr) = 0,
\end{equation}
where $h =(\rho +\rho \epsilon +P)/ \rho$ is the fluid specific enthalpy,
$\gamma$ is the Lorentz factor between the fluid and co-orbiting observers,
and $\gamma_0$ is the Lorentz factor between the co-orbiting and Eulerian
observers. The quantity $\Psi$ is the fluid velocity potential, which 
for an irrotational fluid and an Eulerian observer is related to
the fluid 3-velocity $U^i$ as
\begin{equation}
  U^i = \frac{\psi^{-4}}{\alpha u^t h} \tilde{\gamma}^{ij} \partial_j \Psi,
\end{equation}
where $u^t$ is the time component of the fluid 4-velocity 
(see Refs.~\cite{GourgGTMB01,TanigS10,Gourgoulhon12} for more details).

We model neutron stars with realistic equations of state (EOS),
using piecewise polytrope segments as introduced by
Read {\it et al.}~\cite{ReadLOF09,ReadMSUCF09}. 
In particular, we set the number of polytrope segments to four and
choose the model of APR4 and H4 in Ref.~\cite{ReadLOF09}.
We remind readers that the APR4 EOS~\cite{AkmalPR98} is derived
by a variational method with modern nuclear potentials among neutrons,
protons, electrons, and muons.
The H4 EOS~\cite{LackeNO06} is derived by a relativistic mean-field
theory and includes effects of hyperons. We set the internal flow
in the neutron star to be irrotational as seen by an inertial
observer at infinity~\cite{Kocha92}.

There are two free parameters in the DEF model: 
(i) $B$, which appears in $\omega$,
and (ii) $\varphi_0$ the value of $\varphi$ at spatial infinity.
As discussed in Ref.~\cite{ShibaTOB14}, we choose those values,
taking into account the observational constraints
from neutron-star$-$white-dwarf binaries~\cite{Freir12,Anton13,BhatBV08}. 
We set $\varphi_0 =1 \times 10^{-5}$ and vary $B$ from 8.0 to 9.0
for APR4 EOS, while $\varphi_0=5 \times 10^{-5}$ and $B$
from 8.5 to 9.5 for H4 EOS.

\section{Numerical results} \label{sec3}

We use the spectral-method library \textsc{LORENE}, developed
by the numerical-relativity group at the Observatory of Meudon~\cite{lorene},
and construct a numerical code to compute quasiequilibrium configurations
of binary neutron stars in the DEF model. The code is similar to those
developed in Refs.~\cite{GourgGTMB01,TanigG02,TanigG03,TanigS10} 
for binary neutron stars in general relativity, in particular, 
in Ref.~\cite{TanigS10}.

We consider irrotational, equal-mass binary neutron stars 
whose total mass is $m = 2.7M_{\odot}$ at infinite separation,
and we construct sequences fixing the baryonic rest mass of
the two neutron stars,
\begin{equation}
  M_{\rm B}^{(A)} =\int_{\rm star~A} \rho u^t \sqrt{-g} d^3 x~~~(A=1~{\rm or}~2),
  \label{eq:restmass}
\end{equation}
and varying the orbital separation. As found in Ref.~\cite{ShibaTOB14},
for both APR4 and H4 EOSs spontaneous scalarization does not occur
in each star whose mass in isolation is $1.35 M_{\odot}$. Thus, 
the baryonic rest masses corresponding to the mass of
$1.35 M_{\odot}$ in the scalar-tensor case are $1.50 M_{\odot}$ for APR4
and $1.47 M_{\odot}$ for H4, respectively. Those values are, basically,
the same as those in general relativity. 
Note that the total mass $m$ is the asymptotic value of
the tensor mass defined by $M_{\rm T}=M_{\rm ADM} +M_{\rm S}$~\cite{Lee74}
at infinite separation. Here $M_{\rm ADM}$ is the Arnowitt-Deser-Misner (ADM)
mass and $M_{\rm S}$ is the scalar mass. 
The latter is defined as the monopole part of the scalar field, $\phi$,
which is expanded as $\phi = \phi_0 +2M_{\rm S}/r +{\cal O}(1/r^2)$
for $r \rightarrow \infty$ where $r$ is the radial coordinate. 
In Appendix~\ref{app2}, we show a convergence test for the scalar mass,
varying the numerical resolution (i.e., the number of collocation points).

\begin{figure}
  \vspace{0.5cm}
  \includegraphics[width=8.0cm]{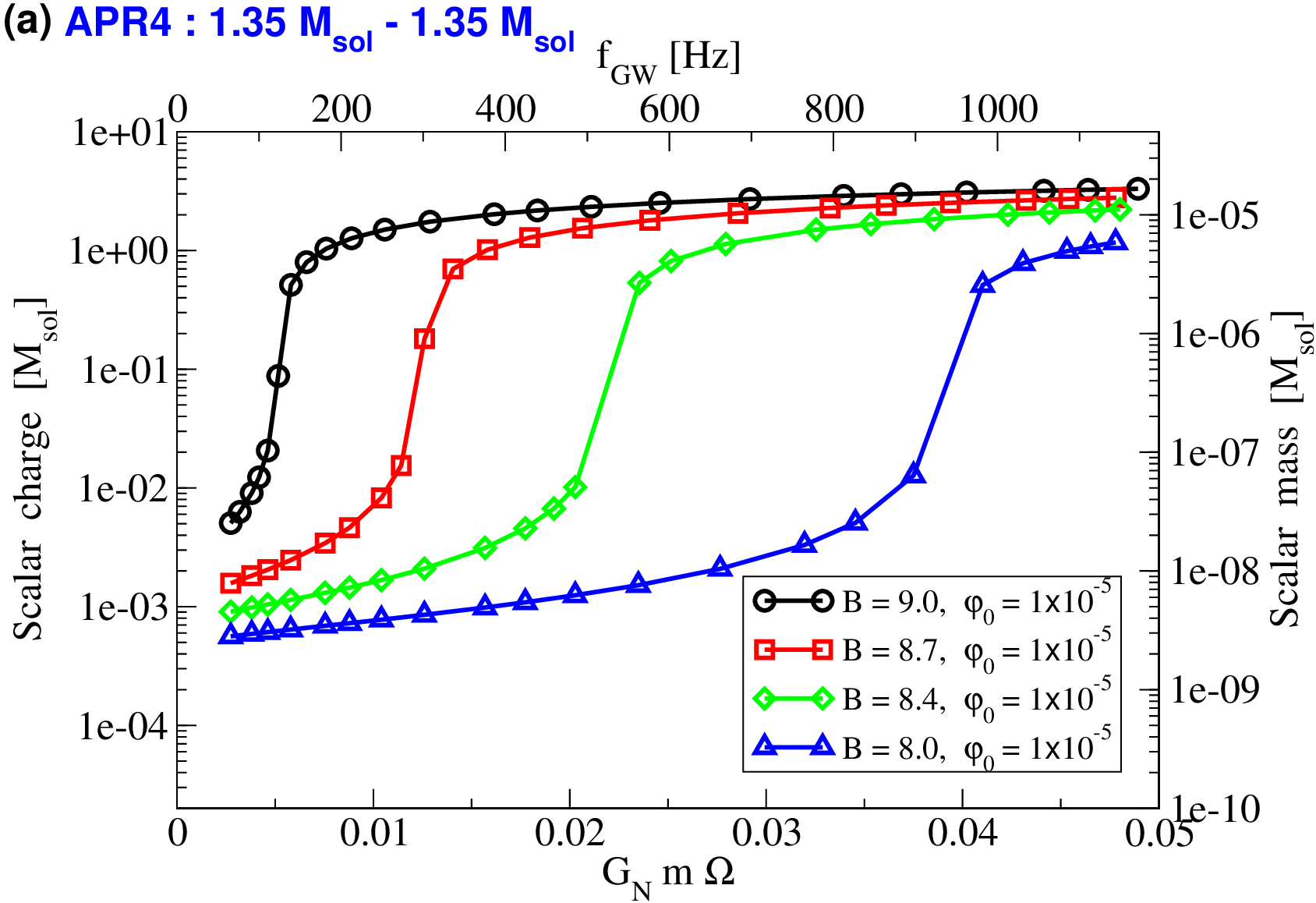} \\
\vspace{0.4cm}
  \includegraphics[width=8.0cm]{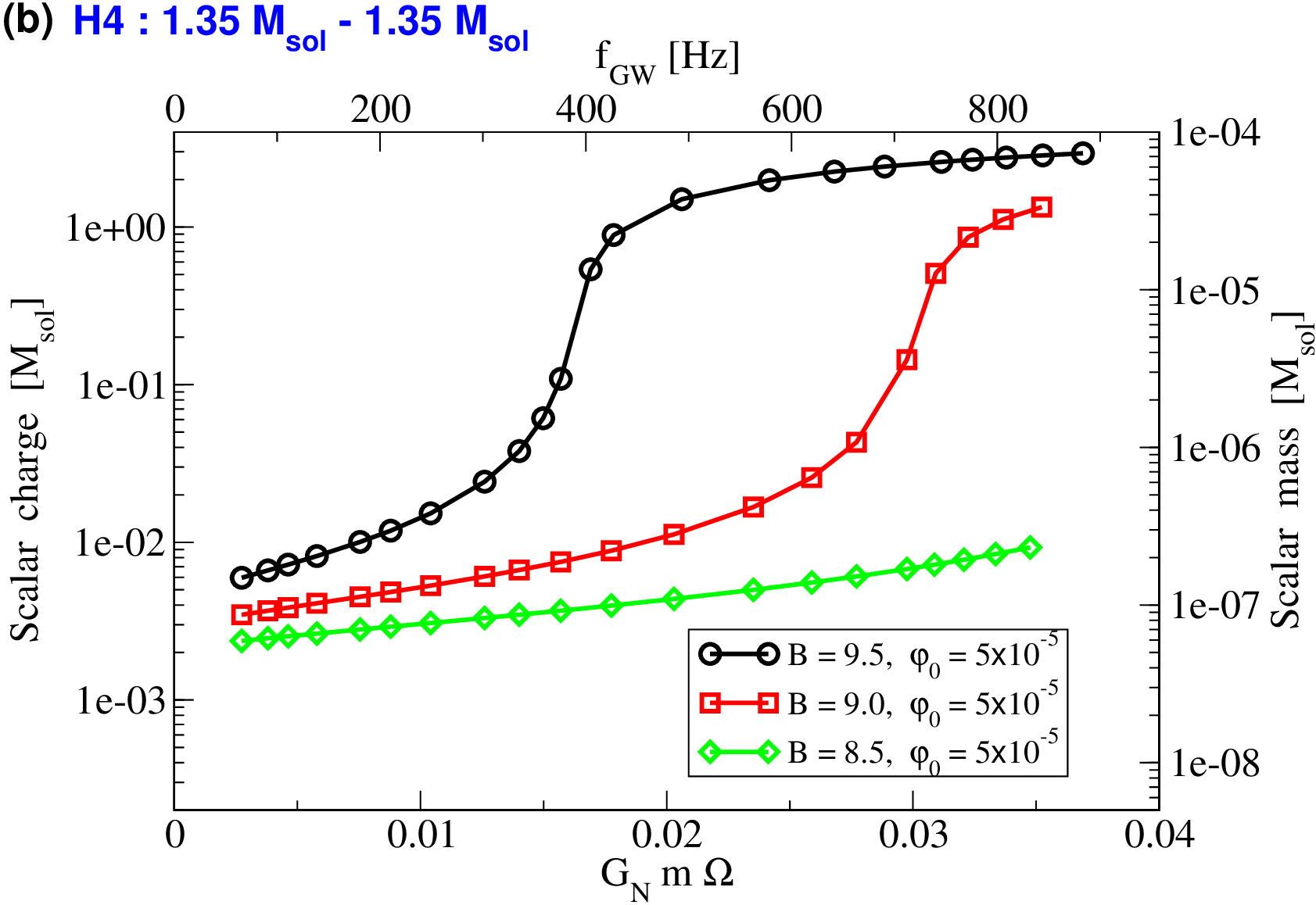}
  \caption{Scalar charge (left $y$ axis) or scalar mass (right $y$ axis)
    as a function of the orbital angular frequency normalized to the
    tensor mass at infinite separation (lower $x$ axis) 
    or as a function of the frequency of GWs 
    defined by $f_{\rm GW} \equiv \Omega/\pi$ from a binary neutron 
    star with $m = 2.7 M_\odot$ (upper $x$ axis).
    Upper panel (a) shows results for APR4 EOS.
    Black solid curve with open circles, red with open squares,
    green with open diamonds, and blue with open triangles are,
    respectively, for the cases $B=9.0$, 8.7, 8.4, and 8.0.
    Lower panel (b) shows results for H4 EOS.
    Black solid curve with open circles, red with open squares,
    and green with open diamonds are, respectively,
    for the cases $B=9.5$, 9.0, and 8.5.
    \label{fig1}}
\end{figure}

\subsection{Scalar charge and scalar mass} \label{sec:scalar}

In Fig.~\ref{fig1} we plot the scalar charge (and scalar mass)
versus the orbital angular frequency for several choices of
the parameters $B$ and $\varphi_0$. 
Here the scalar charge, $M_{\varphi}$, is defined as the monopole
part of the field, $\varphi$, which is expanded as
$\varphi = \varphi_0 +M_{\varphi}/r +{\cal O}(1/r^2)$
for $r \rightarrow \infty$. (Note again that in this paper we employ the 
geometrical units $c=G=1$.) The relation between the scalar charge and the scalar mass is
given by $M_{\varphi}=2M_{\rm S}/(\phi_0 \varphi_0)$.\footnote{Because
the deviation of the quantity $\phi_0 =\exp(\varphi_0^2/2)$ from unity
is on the order of $10^{-9} - 10^{-10}$ as we set
$\varphi_0 =1 \times 10^{-5}$ or $\varphi_0 =5 \times 10^{-5}$,
the relation between the scalar charge and the scalar mass is
approximately given by $M_{\varphi} \simeq 2M_{\rm S}/\varphi_0$
(see Ref.~\cite{ShibaTOB14} for more details).}
We clearly see the onset of dynamical scalarization as the
orbital separation (angular frequency) decreases (increases),
except for the case of H4 EOS $B=8.5$. Those results confirm what is found
when evolving in full numerical relativity a binary neutron star
with APR4 and H4 EOSs~\cite{ShibaTOB14}. In particular, the onset of
dynamical scalarization was well captured by the scalarization condition
derived in Sec. III~B of Ref.~\cite{ShibaTOB14}. For example, Table I of
Ref.~\cite{ShibaTOB14} predicted that the dynamical scalarization for
a binary system of $(1.35 + 1.35) M_{\odot}$ in the case of APR4 $B=9.0$
sets in at around the orbital separation of
$a \simeq 91 M_{\odot} \simeq 134 ~{\rm km}$.
If we regard this separation as the coordinate separation of
our present computation, it corresponds to the orbital angular
frequency of $G_{\rm N} m \Omega \simeq 0.005$.
This is confirmed by Fig.~\ref{fig1} where dynamical scalarization
clearly occurs at around the orbital angular frequency of
$G_{\rm N} m \Omega \simeq 0.005$.

\begin{table}
\caption{We list the orbital angular frequency and GW 
frequency ($f_{\rm GW} \equiv \Omega/\pi$) 
at the onset of dynamical scalarization.
\label{table1}}
\begin{ruledtabular}
\begin{tabular}{lcc}
Models & $G_{\rm N} m \Omega_{\rm dyn\mbox{-}scal}$
& $f_{\rm GW,dyn\mbox{-}scal}~[{\rm Hz}]$ \\
APR4 (9.0) & 0.0051 & 123 \\
APR4 (8.7) & 0.0125 & 298 \\
APR4 (8.4) & 0.0223 & 534 \\
APR4 (8.0) & 0.0395 & 946 \\
H4 (9.5)   & 0.0163 & 391 \\
H4 (9.0)   & 0.0302 & 724 \\
\end{tabular}
\end{ruledtabular}
\end{table}

In Table~\ref{table1} we show the orbital angular frequencies 
and GW frequencies at the onset of dynamical scalarization.
These quantities are extracted from Fig.~\ref{fig1} using the
fit to the scalar charge obtained in Appendix~\ref{app3}
[see Eq.~(\ref{eq:scfit})]. (In particular $G_{\rm N} m \Omega_{\rm dyn\mbox{-}scal}$ 
is the angular orbital frequency at which the fit function intersects 1.)
From Fig.~\ref{fig1} we see that the scalar charge at the orbital
angular frequencies listed in Table~\ref{table1} is about $0.1 M_{\odot}$.
This value is only $4 \%$ of the total mass $m$; thus, at the onset of dynamical 
scalarization the effect of the scalar field onto the dynamics is negligible.
However, as dynamical scalarization proceeds, the scalar charge rapidly
increases by one order of magnitude, 
affecting the subsequent evolution of the binary system.

For the cases of APR4 EOS $B=8.0$ and H4 EOS $B=9.0$,
the numerical-relativity simulations carried out in Ref.~\cite{ShibaTOB14}
showed that dynamical scalarization did not occur during the inspiral
but at merger (see Table~II of Ref.~\cite{ShibaTOB14}). 
On the other hand, Fig.~\ref{fig1} predicts that for APR4 EOS $B=8.0$ and 
H4 EOS $B=9.0$ dynamical scalarization occurs before the end of the 
quasiequilibrium sequence. This contradiction may arise because
toward the end of the inspiral the infall velocity becomes large and 
the quasiequilibrium model may lose accuracy.

Dynamical scalarization was first found in Ref.~\cite{BarauPPL13} 
using polytropic EOS and it was further investigated by the same authors
in Ref.~\cite{PalenBPL14} using 2.5PN equations of motion augmented
by a set of equations that phenomenologically describe the increase 
of scalar charge as the two neutron stars come closer to each other.
Figure~3 in Ref.~\cite{PalenBPL14} is similar to our Fig.~\ref{fig1},
although the former is obtained treating binary neutron stars as two,
isolated spherical neutron stars and using 2PN equations of motion
for circular orbits~\cite{MirshW13}, instead of evolving the binary system
along a sequence of quasiequilibrium configurations in the DEF model.

\begin{figure}
  \vspace{0.5cm}
  \includegraphics[width=7.5cm]{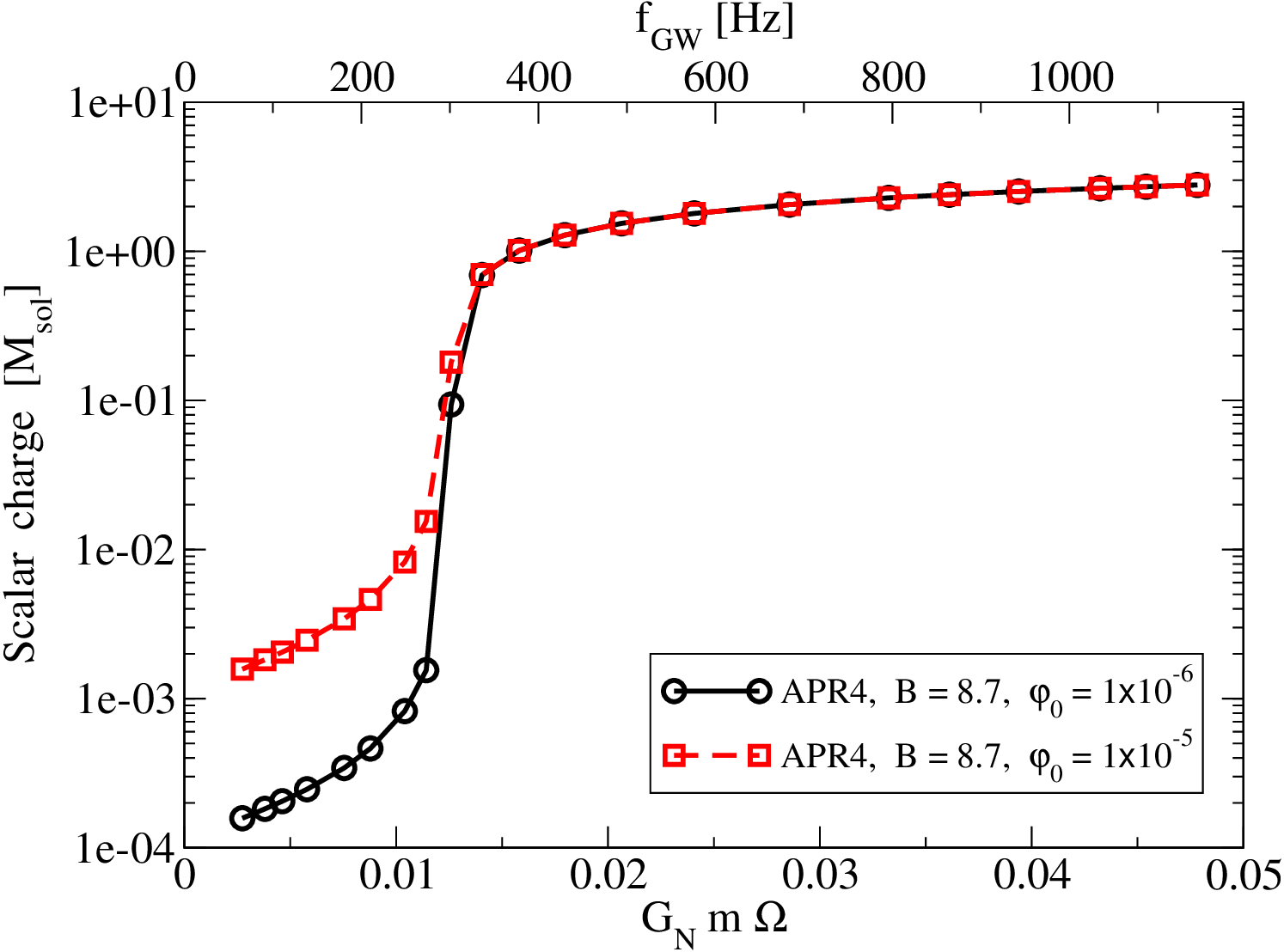}
  \caption{Scalar charge as a function of the orbital angular
    frequency normalized to the tensor mass at infinite separation
    (lower $x$ axis) or as a function of the frequency of GWs (upper $x$ axis).
    Both curves are computed for the case
    of APR4 EOS $B=8.7$, but the value of $\varphi_0$ is set to
    $1 \times 10^{-5}$ (red dashed with open squares) and $1 \times 10^{-6}$
    (black solid with open circles).
    \label{fig2}}
\end{figure}

The results in Fig.~\ref{fig1} are not very sensitive to the value of
$\varphi_0$. Indeed, we show in Fig.~\ref{fig2} that when we decrease
the value of $\varphi_0$ to $1 \times 10^{-6}$ for APR4 EOS, dynamical
scalarization occurs at almost the same orbital angular frequency.
Note that at low orbital frequencies, the scalar charge in the case
$\varphi_0 =1 \times 10^{-6}$ is one order of magnitude smaller than
in the case $\varphi_0 =1 \times 10^{-5}$. This confirms that at large
separation the scalar charge is proportional to $\varphi_0$, as found in
Ref.~\cite{ShibaTOB14}. After dynamical scalarization occurs the scalar
charge has the same value, independently on $\varphi_0$.

Note that the numerically constructed quasiequilibrium sequences
in Fig.~\ref{fig1} (also in Figs.~\ref{fig2}$-$\ref{fig5})
do not end at the onset of mass shedding from the neutron star's surface,
but stop just before it. This is because it is impossible to treat
a cuspy shape within the spectral method. Thus, we are obliged to stop
the computation before the onset of mass shedding
where the neutron star acquires a cuspy shape. 
Note also that in principle Gibbs phenomena could be present 
at the surface of the star even before the mass-shedding limit takes place.
This is due to large differences in the density's derivative. However,  
because the \textsc{LORENE} spectral code adopts a multidomain method and
surface-fitting coordinates, on each domain the physical fields are
smooth functions and Gibbs phenomena
do not appear (see Ref.~\cite{BonazGM98} for detailed explanations).

\subsection{Binding energy and total angular momentum}

\begin{figure}
  \vspace{0.5cm}
  \includegraphics[width=7.7cm]{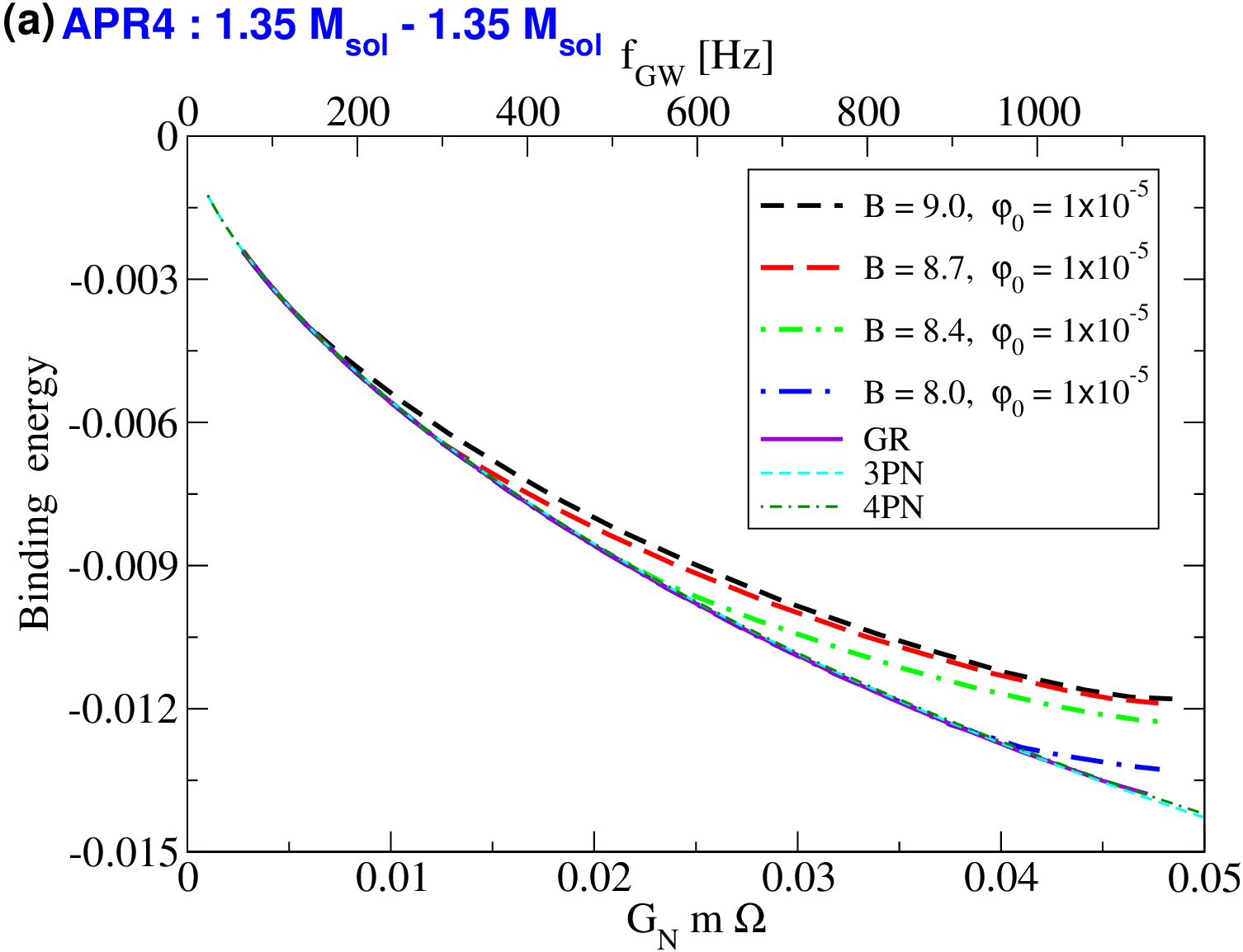} \\
\vspace{0.4cm}
  \includegraphics[width=7.7cm]{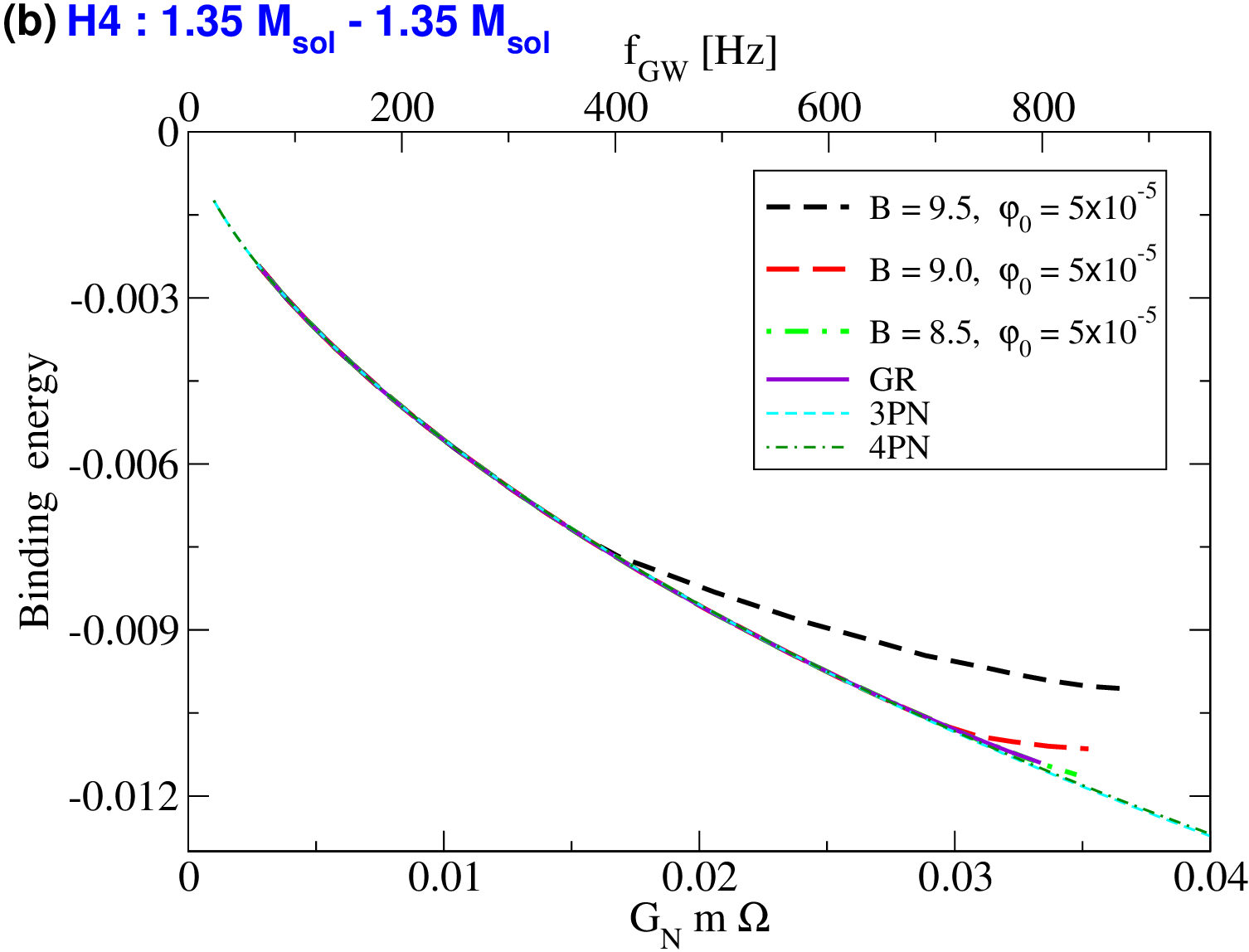}
  \caption{Binding energy as a function of the orbital angular frequency
    normalized to the tensor mass at infinite separation (lower $x$ axis)
    or as a function of the frequency of GWs from a binary
    neutron star with $m = 2.7 M_\odot$ (upper $x$ axis).
    Upper panel (a) shows results for APR4 EOS. Black short-dashed,
    red long-dashed, green dot-short-dashed, and blue dot-long-dashed curves
    are, respectively, the cases: $B=9.0$, 8.7, 8.4, and 8.0.
    Lower panel (b) shows results for H4 EOS. Black short-dashed,
    red long-dashed, and green dot-dashed curves are, respectively,
    the cases: $B=9.5$, 9.0, and 8.5.
    In both panels, the purple solid curve is drawn by using
    a quasiequilibrium sequence in general relativity (GR).
    Cyan dashed and dark-green dot-dashed curves refer to the 
    3PN and 4PN binding energy in general relativity,
    respectively~\cite{BiniD13}
    (see also Refs.~\cite{BarauBT12,JaranS12,JaranS13}).
    \label{fig3}}
\end{figure}

We plot in Fig.~\ref{fig3} the normalized binding energy,
$(M_{\rm T}-m)/m$, along the quasiequilibrium sequences of
binary neutron stars versus the orbital angular frequency.
We find that after the onset of dynamical scalarization the
binding energy in the scalar-tensor case decreases less rapidly
than in general relativity (GR), differing at most by $14 \%$ 
at high frequencies for the cases considered.
This implies that binary neutron stars undergoing dynamical scalarization 
in the DEF theory spiral in more quickly than in general relativity, 
if the amount of energy flux of gravitational radiation
in the scalar-tensor case is equal to the one in general relativity.
As we shall see below, the former can be much larger than the latter,  
so the binary neutron star approaches the merger even more quickly 
once the energy flux in scalar-tensor theory is also taken into 
account.

The binding energy is defined by the difference between the tensor mass
at a given separation, $M_{\rm T}$, and that at infinity, $m$.
The tensor mass is given by the sum of the ADM mass, $M_{\rm ADM}$,
and the scalar mass, $M_{\rm S}$~\cite{Lee74}. Quite interestingly, 
we find that the scalar mass is not responsible of the large increase
of the tensor mass; the latter increases because of the large increase
(three orders of magnitude) of the scalar field in the ADM mass.
As seen in Eq.~(\ref{eq:deltapsi}), the scalar field and its derivatives
enter the source term of the Poisson-like equation of the conformal
factor, which determines the ADM mass.

\begin{figure}
  \vspace{0.5cm}
  \includegraphics[width=7.7cm]{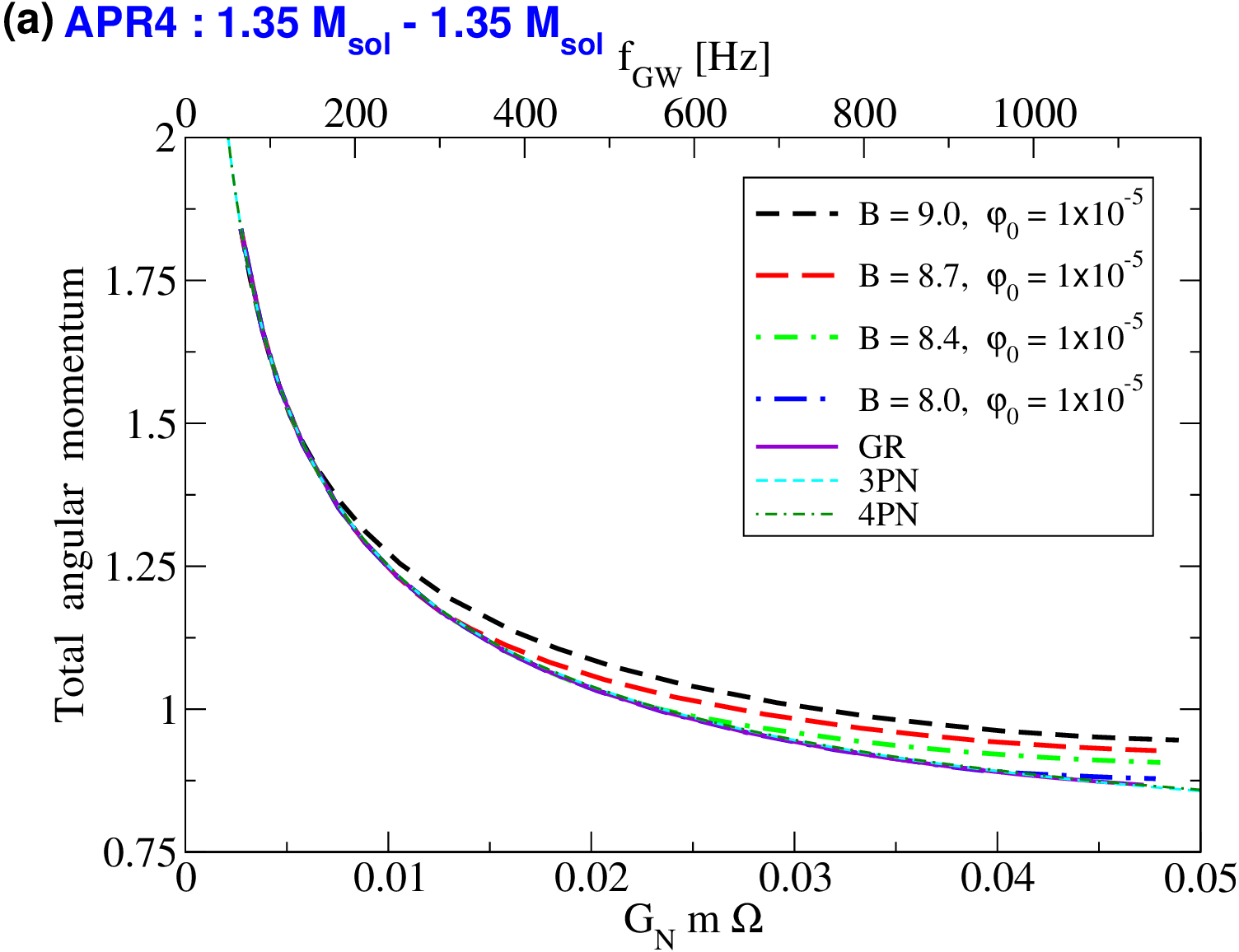} \\
\vspace{0.4cm}
  \includegraphics[width=7.7cm]{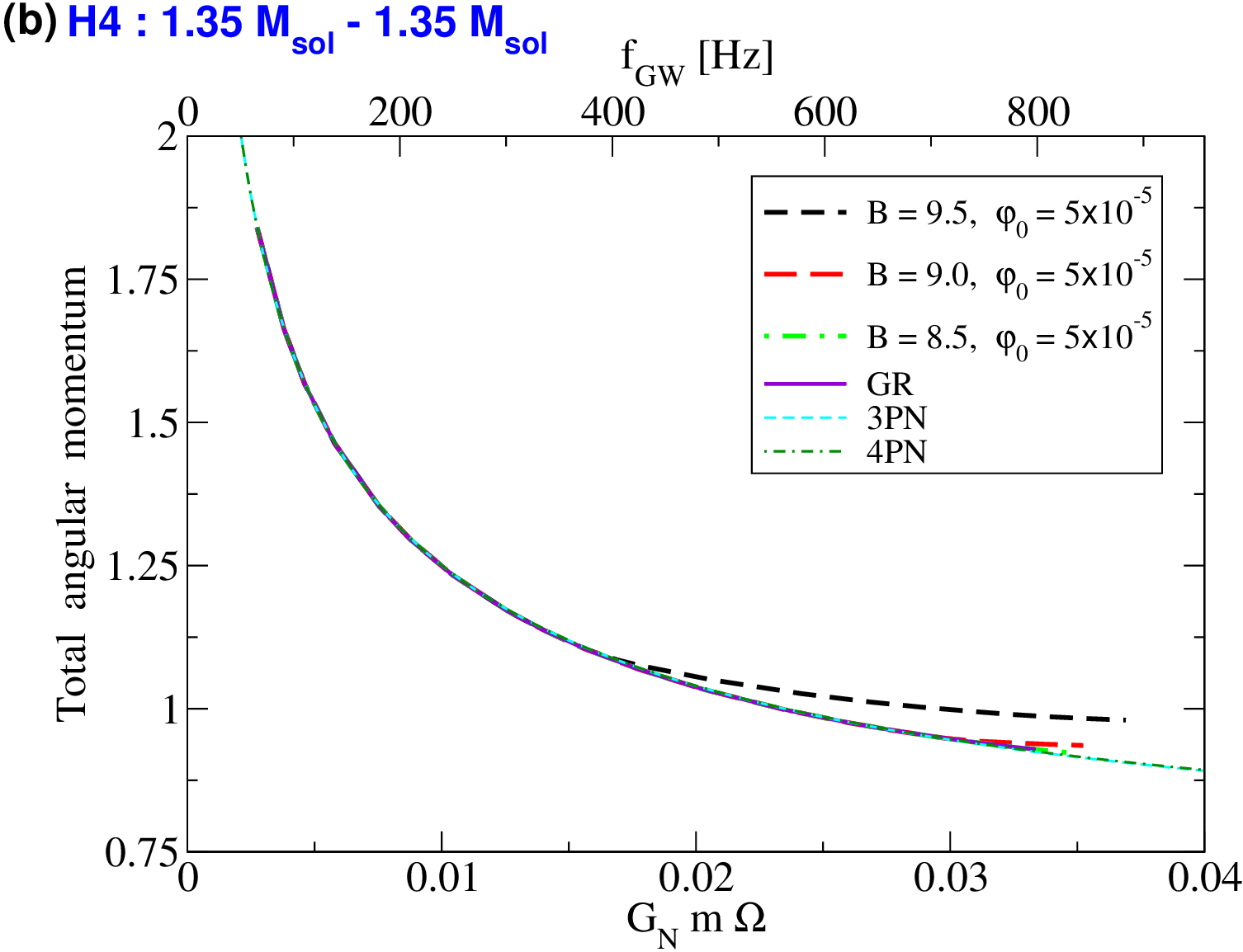}
  \caption{Same as Fig.~\ref{fig3} but for the total angular momentum.
    Cyan dashed and dark-green dot-dashed curves refer to the 
    3PN and 4PN angular momentum in general relativity,
    respectively. Those curves are calculated using 
    Refs.~\cite{LeTieBW12,BiniD13}.
    \label{fig4}}
\end{figure}

In Fig.~\ref{fig4} we plot the total angular momentum,
$J/(G_{\rm N} m^2)$, along the quasiequilibrium sequences
versus the orbital angular frequency.
The behavior of the total angular momentum is basically the same as
the binding energy; i.e., after the onset of dynamical scalarization
the total angular momentum in the scalar-tensor case decreases less rapidly
than in general relativity.

It is worth noticing that in some cases the binding energy/total angular
momentum may reach their minimum before the onset of mass shedding. 
The sequences shown in Figs.~\ref{fig3} and \ref{fig4}
terminate slightly before the onset of mass-shedding
because it is impossible to treat a cuspy shape within 
the spectral method, as we mentioned at the end of Sec.~\ref{sec:scalar}.
In order to calculate at which orbital angular frequency the sequences
encounter the mass-shedding point, we compute the sensitive mass-shedding
indicator $\chi$,
\begin{equation}
  \chi \equiv \frac{(\partial (\ln h)/ \partial r)_{\rm eq}}
  {(\partial (\ln h)/ \partial r)_{\rm pole}}
\end{equation}
as a function of the orbital angular frequency.
The above quantity is the ratio between the radial derivative of the enthalpy
computed in the equatorial plane at the surface along the direction toward
the companion star and the one at the surface of the pole of the star.
The indicator takes the value $\chi=1$ for spherical stars and it is  
$\chi=0$ in the mass-shedding limit. We extrapolate the sequences of 
$\chi$ to the mass-shedding limit and determine the orbital angular 
frequency at that point. (For more details on this method see Sec.~4.3 
in Ref.~\cite{TanigS10}.) After determining the orbital angular frequency 
at the onset of mass shedding, we then extrapolate the binding energy curve to
that frequency.

By this procedure we find that the sequences of APR4 EOS $B=9.0$,
8.7, 8.4, 8.0, and H4 EOS $B=9.5$, $B=9.0$ reach the minimum of
the binding energy (i.e., the onset of secular instability~\cite{FriedUS02})
before the mass shedding. Thus, in these cases the binary neutron stars
terminate the quasiequilibrium sequence at that point, and plunge.
By contrast the sequence of H4 EOS $B=8.5$ terminates at the onset
of mass shedding.

\subsection{Central baryonic rest-mass density}

\begin{figure}
  \vspace{0.5cm}
  \includegraphics[width=7.5cm]{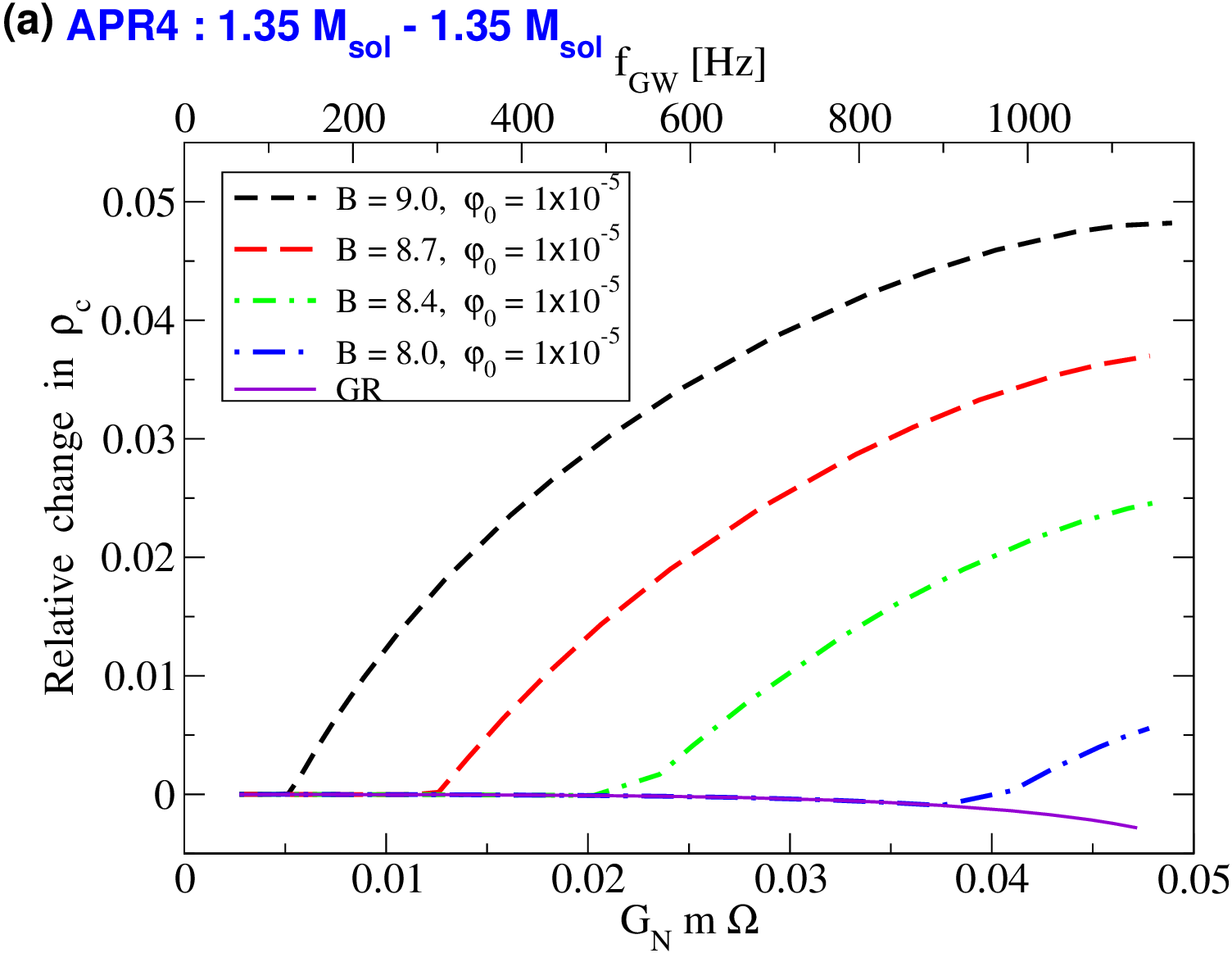} \\
\vspace{0.4cm}
  \includegraphics[width=7.5cm]{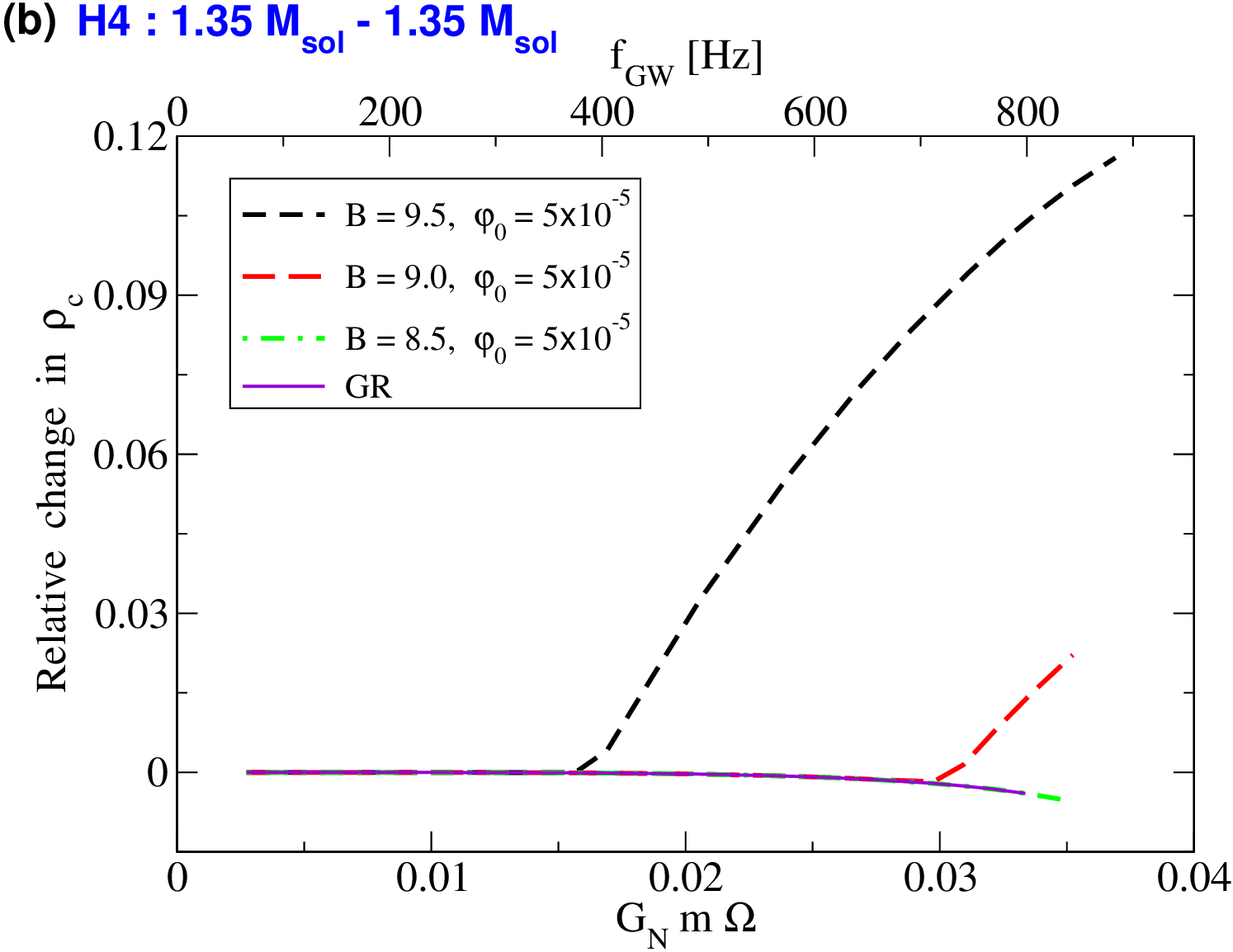}
  \caption{Same as Fig.~\ref{fig3} but for the relative change
    in the central baryonic rest-mass density of a neutron star.
    \label{fig5}}
\end{figure}

In general relativity the central baryonic rest-mass density
of a neutron star in irrotational binary systems always decreases
as the orbital frequency increases (see, e.g., Ref.~\cite{TanigG02}).
We find that this is not the case in the scalar-tensor model under
investigation. After the onset of dynamical scalarization, we find that
the central baryonic rest-mass density starts increasing
(instead of continuing decreasing)
as the orbital frequency increases, as shown in Fig.~\ref{fig5}.
It is not easy to physically explain this behavior because of nonlinear
effects related to the deformation of the star and the distribution
profiles of metric and matter quantities in the star.
In our computation, the baryonic rest mass of each star defined by 
Eq.~(\ref{eq:restmass}) is fixed along the quasiequilibrium sequences.
To do so, the central value of the baryonic rest-mass density
of a neutron star necessarily increases after dynamical scalarization,
because the remaining part of Eq.~(\ref{eq:restmass}) after dropping
the rest-mass density,
\begin{equation}
  \int_{\rm star~A} u^t \sqrt{-g} d^3x,
\end{equation}
decreases rapidly after dynamical scalarization sets in along a constant
baryonic rest-mass sequence, while it slightly increases before the
dynamical scalarization.

\subsection{Evolution of the orbital angular frequency} 
\label{sec:evolorb}

\begin{figure}
  \vspace{0.5cm}
  \includegraphics[width=7.7cm]{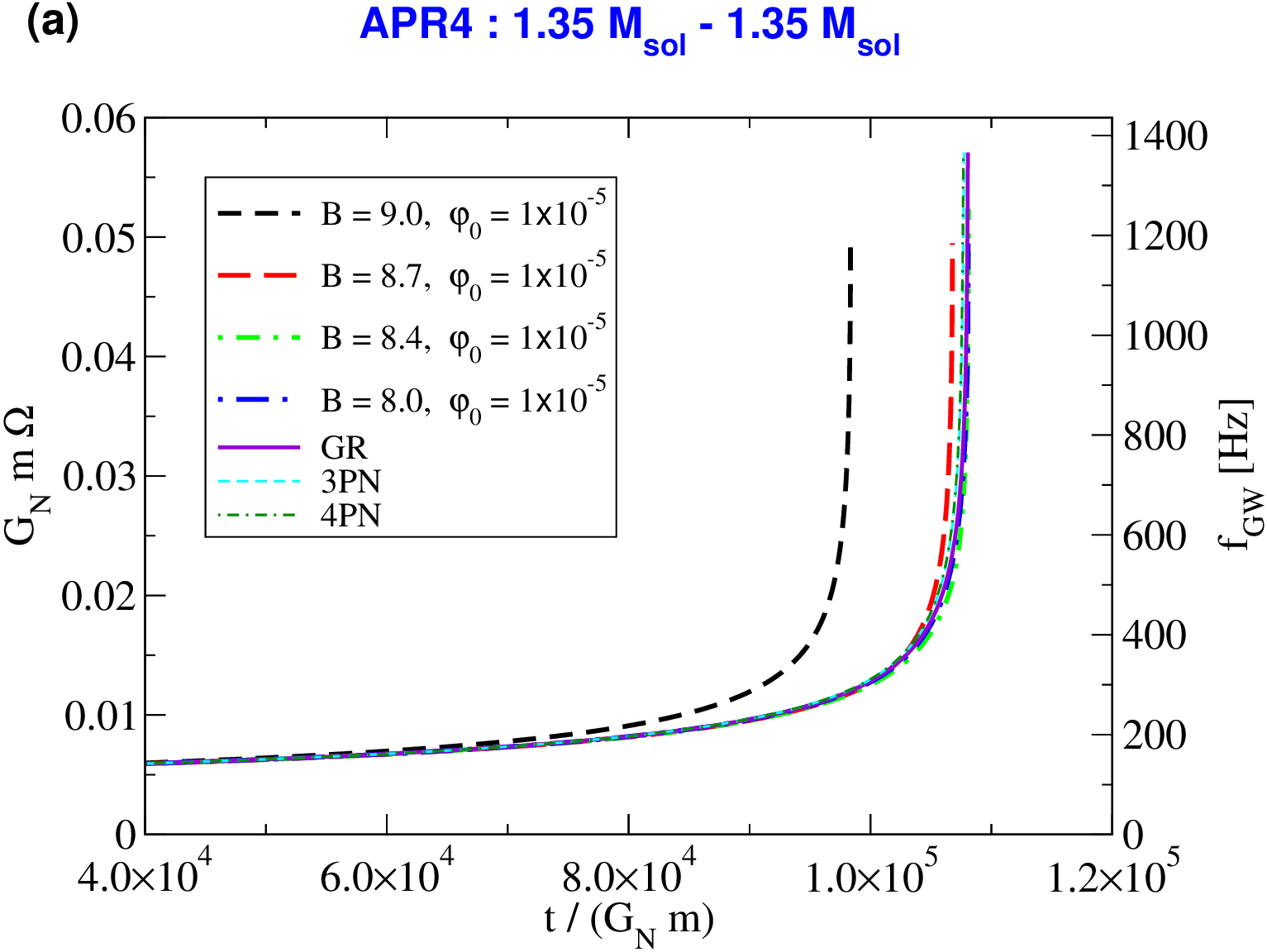} \\
\vspace{0.4cm}
  \includegraphics[width=7.7cm]{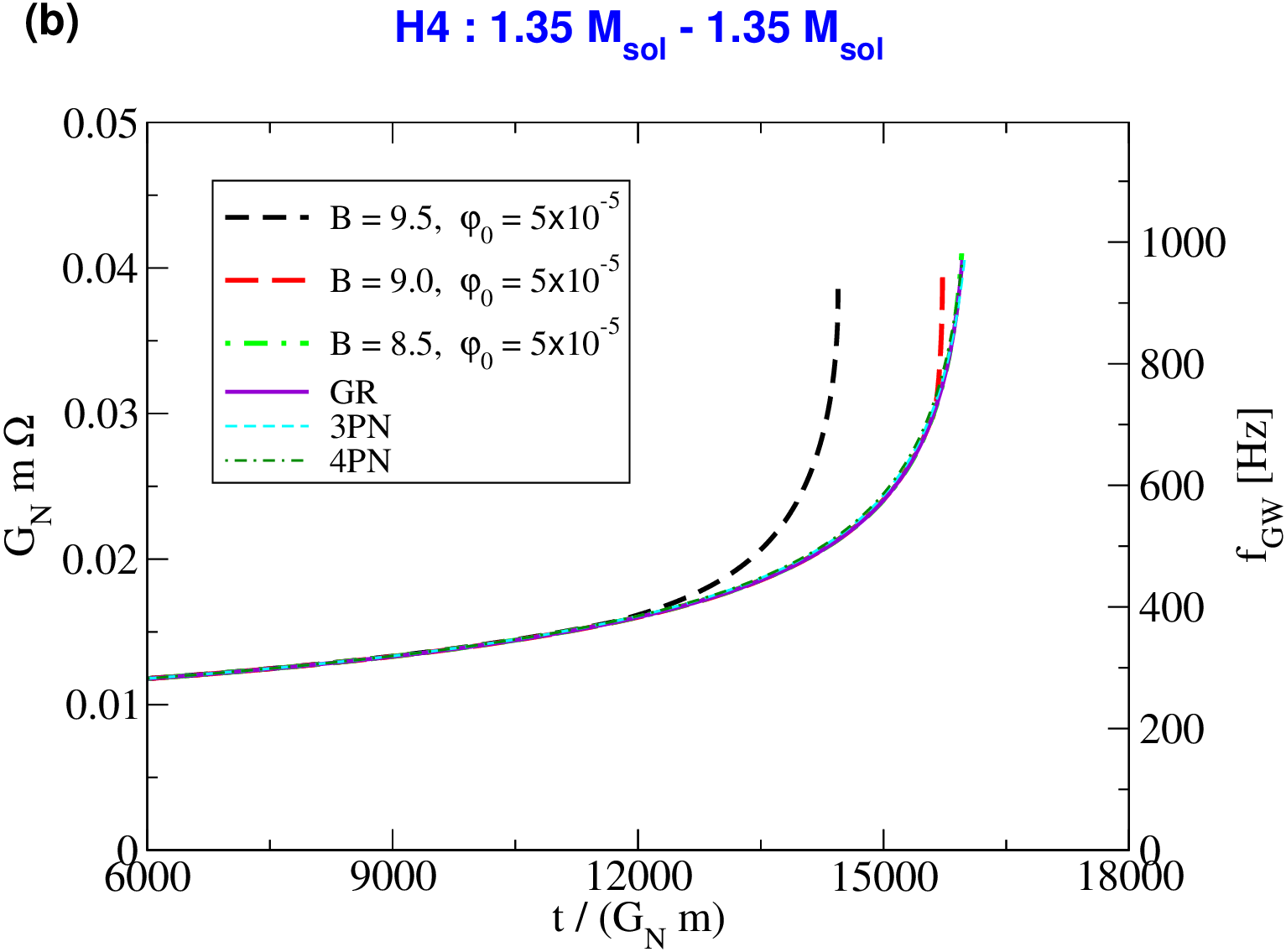}
  \caption{Same as Fig.~\ref{fig3} but for the evolution of
    the orbital angular frequency (left $y$ axis) 
    or the frequency of GWs (right $y$ axis) 
    from a binary neutron star with $m = 2.7 M_\odot$.
    The gravitational energy flux is the choice (i).
    \label{fig6}}
\end{figure}

We want to estimate the orbital angular frequency and GW cycles 
in the DEF model assuming a quasistationary adiabatic evolution.
We follow what was done in Ref.~\cite{TanigS10}. Basically, we use the
balance equation $dE/dt = - {\cal F}$ and integrate
$d \Omega/dt = -{\cal F}/(dE/d\Omega)$. (Note that this PN approximant is
similar to the TaylorT1 approximant in Ref.~\cite{BoyleBKMPSCT07}.)
For $E$ we use the binding energy [fitted to a polynomial in
$x \equiv (G_{\rm N} m \Omega)^{2/3}$] along the quasiequilibrium sequences.
The choice of the energy flux ${\cal F}$ is less straightforward
because we do not know it exactly, so we have to rely on PN
calculations~\cite{DamourEF92,DamourEF96b,DamourEF98,WillZ89,MirshW13,Lang14}.

Since in this paper we consider only the case of equal-mass binaries on
a circular orbit, the monopole and dipole components of the gravitational
radiation vanish. Thus the leading term is the quadrupole component. 
Using Ref.~\cite{DamourEF92} [see in particular Eqs.~(6.40) and (6.41) therein]
we find that the ratio of the quadrupole component generated directly
by the scalar field, ${\cal F}_{\varphi}^{\rm Quadrupole}$, 
and the quadrupole component of the gravitational field,
${\cal F}_{G}^{\rm Quadrupole}$, is given at leading order in the PN
expansion by\footnote{Here and in the following we assume that
the energy flux derived in PN theory expanding the neutron-star masses about
the asymptotic value $\varphi_0$ continues to be valid 
also in the presence of spontaneous and/or dynamical scalarization.}
\begin{equation}
  \frac{{\cal F}_{\varphi}^{\rm Quadrupole}}{{\cal F}_{G}^{\rm Quadrupole}} 
= \frac{\alpha_{\varphi}^2}{6}
  = \frac{1}{6 B} \Bigl( \frac{M_{\varphi, {\rm NS}}}{m_{\rm NS}} \Bigr)^2,
  \label{eq:ratioflux}
\end{equation}
where $M_{\varphi, {\rm NS}}$ is the scalar charge of a neutron star,
and $m_{\rm NS}$ is the tensor mass of a {\it spherical} neutron star, 
i.e., $m_{\rm NS} = m/2 = 1.35 M_{\odot}$. The quantity 
$\alpha_{\varphi}$ is an auxiliary quantity~\cite{DamourEF92,DamourEF93} 
defined by
\begin{equation}
  \alpha_{\varphi} = -\frac{M_{\varphi, {\rm NS}, {\rm DEF}}}{m_{\rm NS}}
  = -\frac{1}{\sqrt{B}} \frac{M_{\varphi, {\rm NS}}}{m_{\rm NS}}. \label{eq:alpha}
\end{equation}
Here $M_{\varphi, {\rm NS}, {\rm DEF}}$ is the scalar charge of a neutron star
defined as the monopole part of the field, $\varphi_{\rm DEF}$,
which is expanded as
$\varphi_{\rm DEF} = \varphi_{0, {\rm DEF}} +M_{\varphi, {\rm NS}, {\rm DEF}}/r +{\cal O}(1/r^2)$
for $r \rightarrow \infty$. (Note that the relation between the var-type
scalar field in this paper, $\varphi$, and that in Ref.~\cite{DamourEF92},
$\varphi_{\rm DEF}$, is given by $\varphi = \sqrt{B} \varphi_{\rm DEF}$.)

Because the scalar charge of a binary system is defined as
a global quantity, we do not know the scalar charge
of the individual stars in the binary system, $M_{\varphi, {\rm NS}}$.
As an approximation, we simply take $M_{\varphi, {\rm NS}} \simeq M_{\varphi}/2$
for the current estimate. From Fig.~\ref{fig1}, we find that the ratio
Eq.~(\ref{eq:ratioflux}) takes the maximum of about 0.028 in the case
of APR4 $B=9.0$ at the end of the sequence.
(The scalar charge at that point is about $M_{\varphi} = 3.3 M_{\odot}$.)
For the other cases the ratio is less than 0.028 throughout
the quasiequilibrium sequences. Thus, since
${\cal F}_{\varphi}^{\rm Quadrupole}$ is at most $3 \%$ of the quadrupole component 
of the gravitational field, ${\cal F}_{G}^{\rm Quadrupole}$, we neglect it.

Considering the above, we make the following choices for the gravitational
energy flux ${\cal F}$ in the balance equation: (i) the 3.5PN flux
(also as a polynomial in $x$) computed in general relativity~\cite{Blanchet06},
and (ii) the quadrupole component of the gravitational field in
the scalar-tensor DEF model, i.e., ${\cal F}_{G}^{\rm Quadrupole} =
(32 \nu^2/5) (G_{\rm N} m (1 + \alpha_{\varphi}^2) \Omega)^{10/3}$ where 
$\nu = m_{\rm NS}^2/m^2$~\cite{DamourEF92,PalenBPL14}.
The choice (i) will allow us to isolate the contribution to the orbital
angular frequency (and the number of GW cycles, see below) due to
the binding energy computed in this paper, while the choice 
(ii) is an estimate of the orbital angular frequency when also the
gravitational radiation in scalar-tensor theory is included.
(Note that today the energy flux in scalar-tensor theory is known only through 
leading quadrupole order.) In Appendix~\ref{app3}, we further discuss
those choices and explain how we derive the energy flux in case (ii) using the 
scalar charge computed in this paper.\footnote{We are currently working
on long-term evolutions of
binary neutron stars extending our previous study~\cite{ShibaTOB14}. 
Preliminary analysis shows that the GW energy flux in the scalar-tensor DEF
model, at a given orbital frequency, is indeed larger than in the
general-relativity case. However, case (ii) seems to overestimate it
and the exact GW luminosity is likely to lie between cases (i) and (ii).}

\begin{figure}
  \vspace{0.5cm}
  \includegraphics[width=7.7cm]{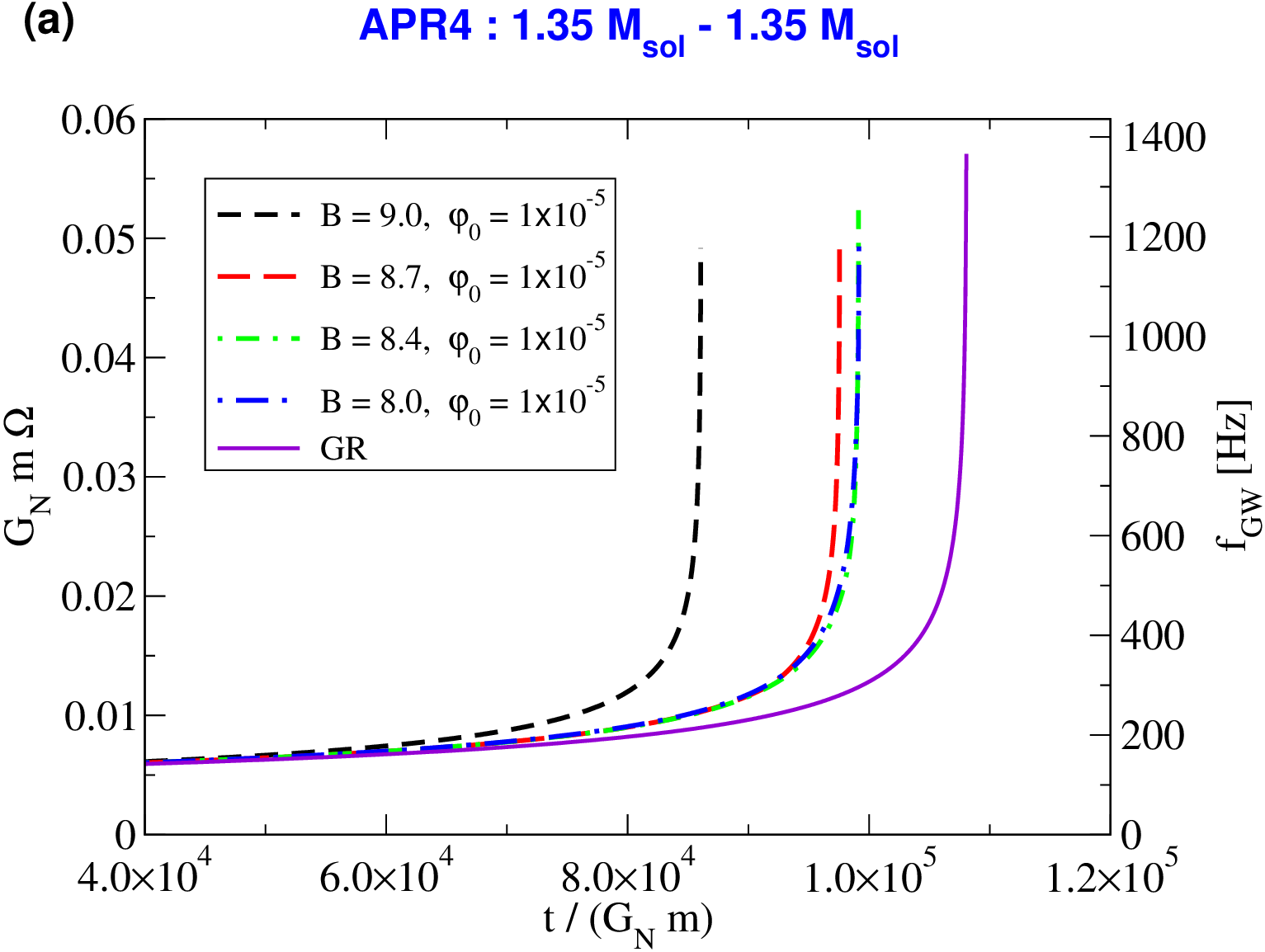} \\
\vspace{0.4cm}
  \includegraphics[width=7.7cm]{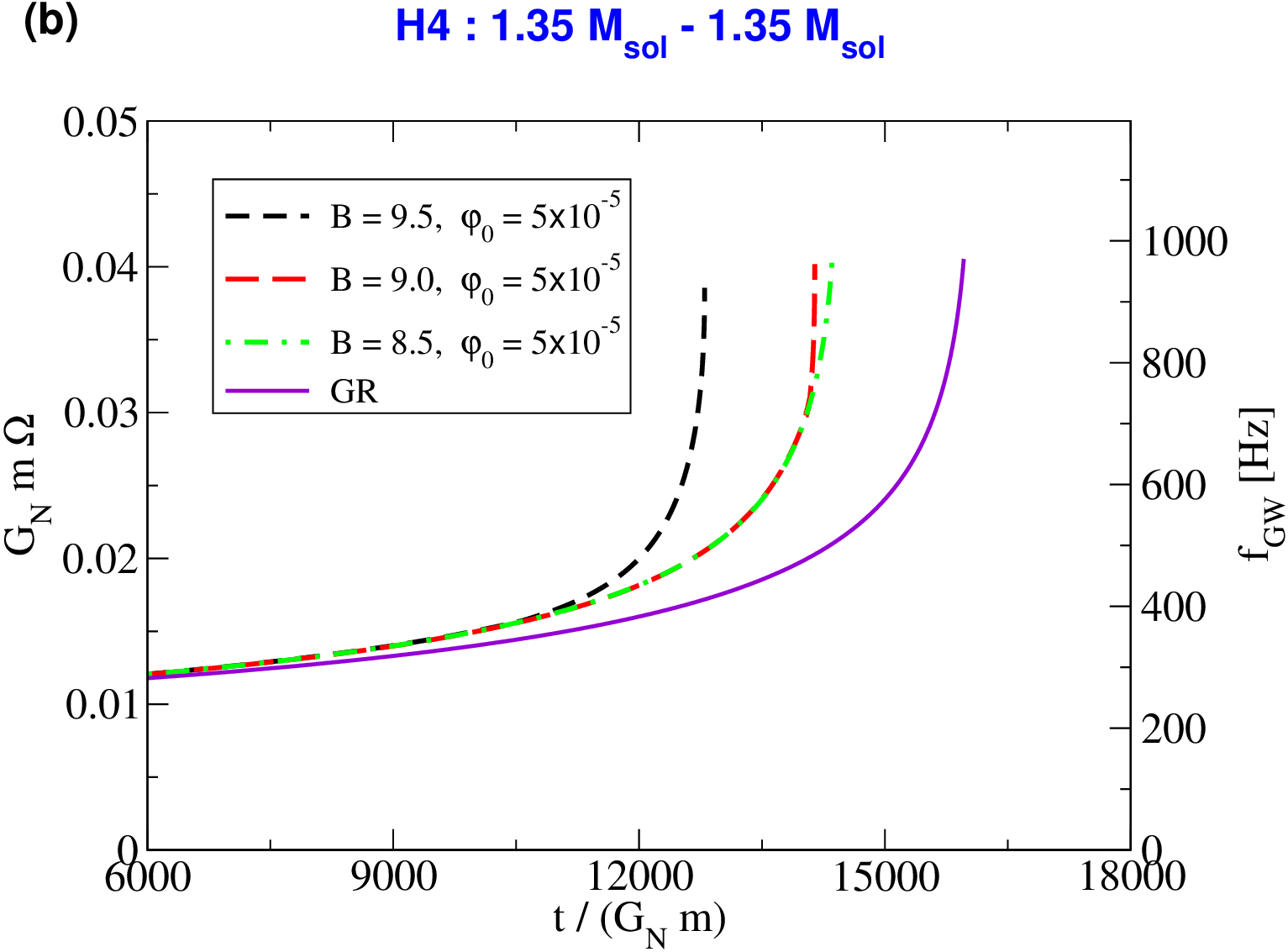}
  \caption{Same as Fig.~\ref{fig6} but for the choice (ii)
    for the gravitational energy flux.
    In the general-relativity case, the 3.5PN energy flux computed
    in general relativity is used.
    These are the same data as used in Fig.~\ref{fig6} (shown as ``GR'').
    \label{fig7}}
\end{figure}

In the examples shown in Fig.~\ref{fig6} for the choice (i) and in
Fig.~\ref{fig7} for the choice (ii),
we set the initial orbital angular frequency to $G_{\rm N} m \Omega =0.005$
for the case of APR4 EOS
and $G_{\rm N} m \Omega =0.01$ for H4 EOS.
The final orbital angular frequency of
each curve corresponds to the end point of the quasiequilibrium sequences.
As explained above when discussing Fig.~\ref{fig3}, the sequences
may have the minimum of the binding energy before the mass-shedding point.
If the minimum is found, we use the orbital angular frequency 
at that point, $G_{\rm N} m \Omega_{\rm ener\mbox{-}min}$, as the final one.
If not, we adopt the orbital angular frequency at the mass-shedding limit,
$G_{\rm N} m \Omega_{\rm mass\mbox{-}shed}$, as the final one.
In Table~\ref{table2} we show the orbital angular frequency at the end point
of each model which is used for the final orbital angular frequency.
For 3PN and 4PN cases, we use the same final orbital angular frequency
as that in general relativity.
Note that the final orbital angular frequency listed in Table~\ref{table2}
is about $G_{\rm N} m \Omega = 0.05$ for APR4 EOS and 0.04 for H4 EOS. 
These values correspond to the frequency of GWs of $\simeq 1200~{\rm Hz}$
and $\simeq 960~{\rm Hz}$ which lie in the high-frequency portion of
the LIGO/Virgo/KAGRA bandwidth. 
As a consequence, the effects discussed in Secs.~\ref{sec:evolorb} and
\ref{sec:gwcycles} may not be observable if the broadband noise 
spectral density is employed. By contrast interferometer configurations
optimized at high frequency may allow us to measure these effects.

It is relevant to point out that if we consider the case of unequal-mass
binaries, there exists the dipole component of the scalar GWs. 
This contribution has the same sign as that of the quadrupole
component~\cite{DamourEF92} and increases the energy flux of the
scalar GWs. Moreover, if we take into account the infall velocity of the stars
in the binary system, the monopole component arises. This also contributes
to increase the energy flux of scalar GWs~\cite{DamourEF92}.
Thus, since the energy flux increases for unequal-mass binaries,
in these cases the merger may occur at even earlier times than what
we estimated in Fig.~\ref{fig6}.

\subsection{Number of gravitational-wave cycles} \label{sec:gwcycles}

\begin{figure}
  \vspace{0.5cm}
  \includegraphics[width=7.cm]{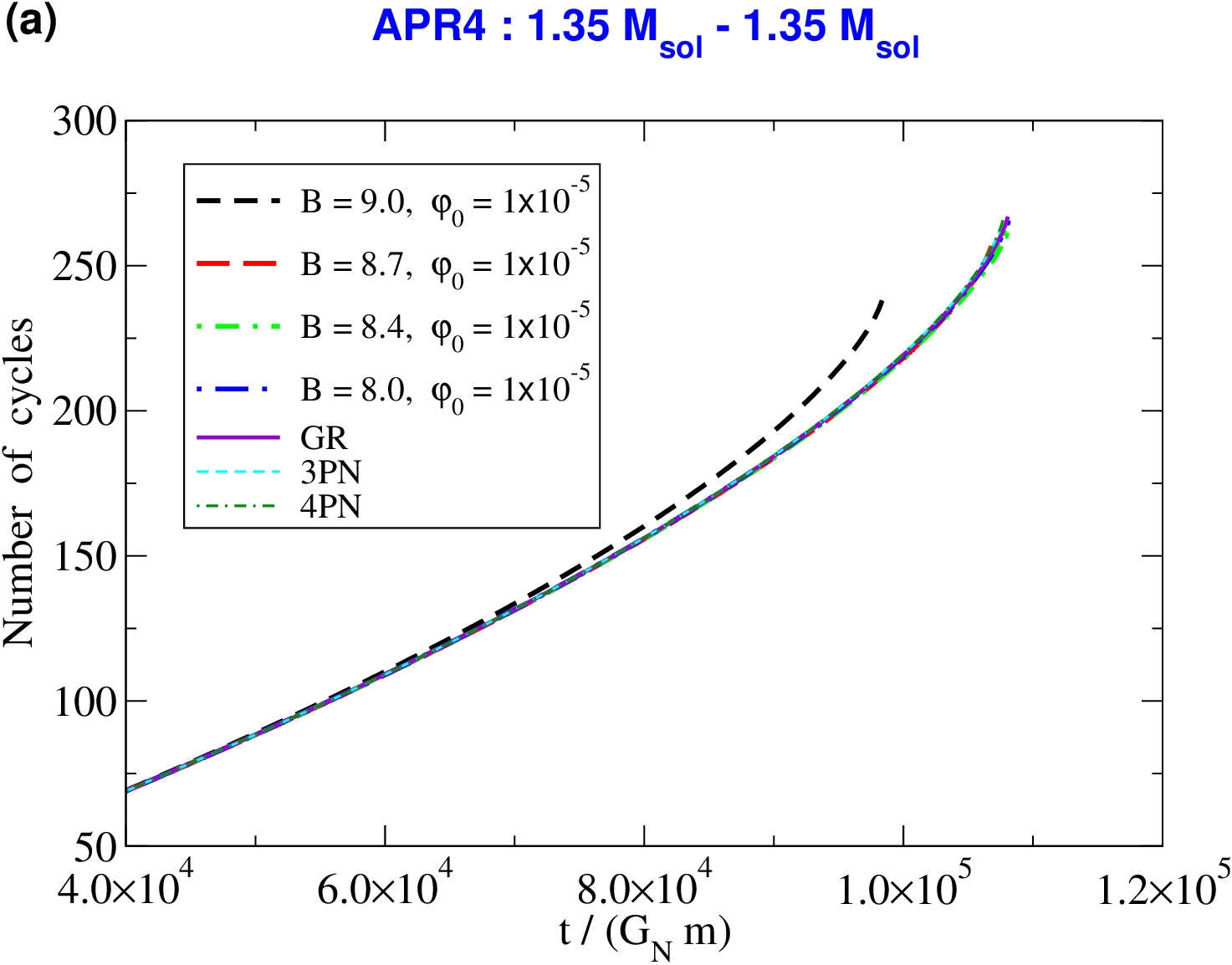} \\
\vspace{0.4cm}
  \includegraphics[width=7.cm]{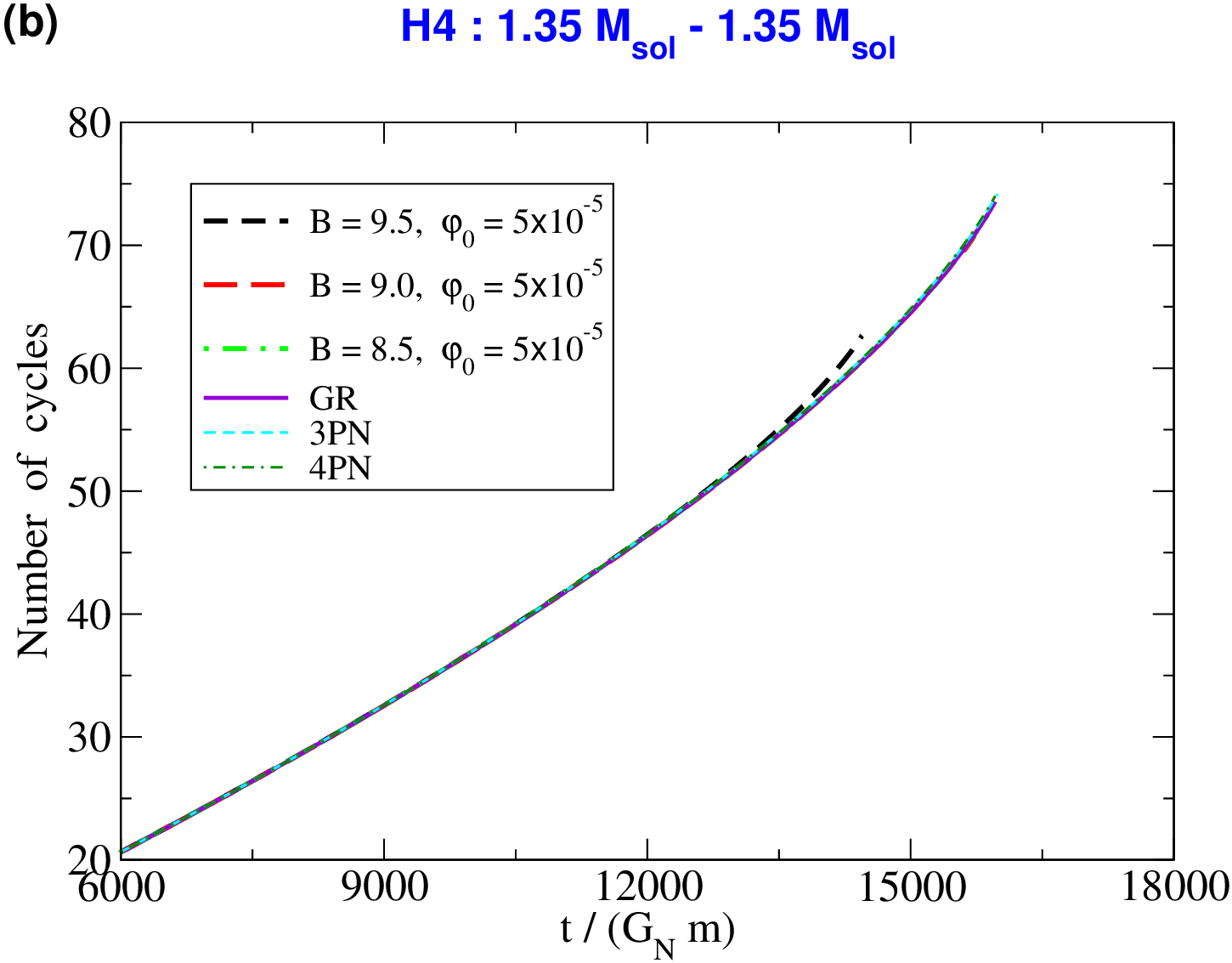}
  \caption{Same as Fig.~\ref{fig3} but for the evolution of
    the number of GW cycles.
    The gravitational energy flux is the choice (i).
    \label{fig8}}
\end{figure}

\begin{figure}
  \vspace{0.5cm}
  \includegraphics[width=7.cm]{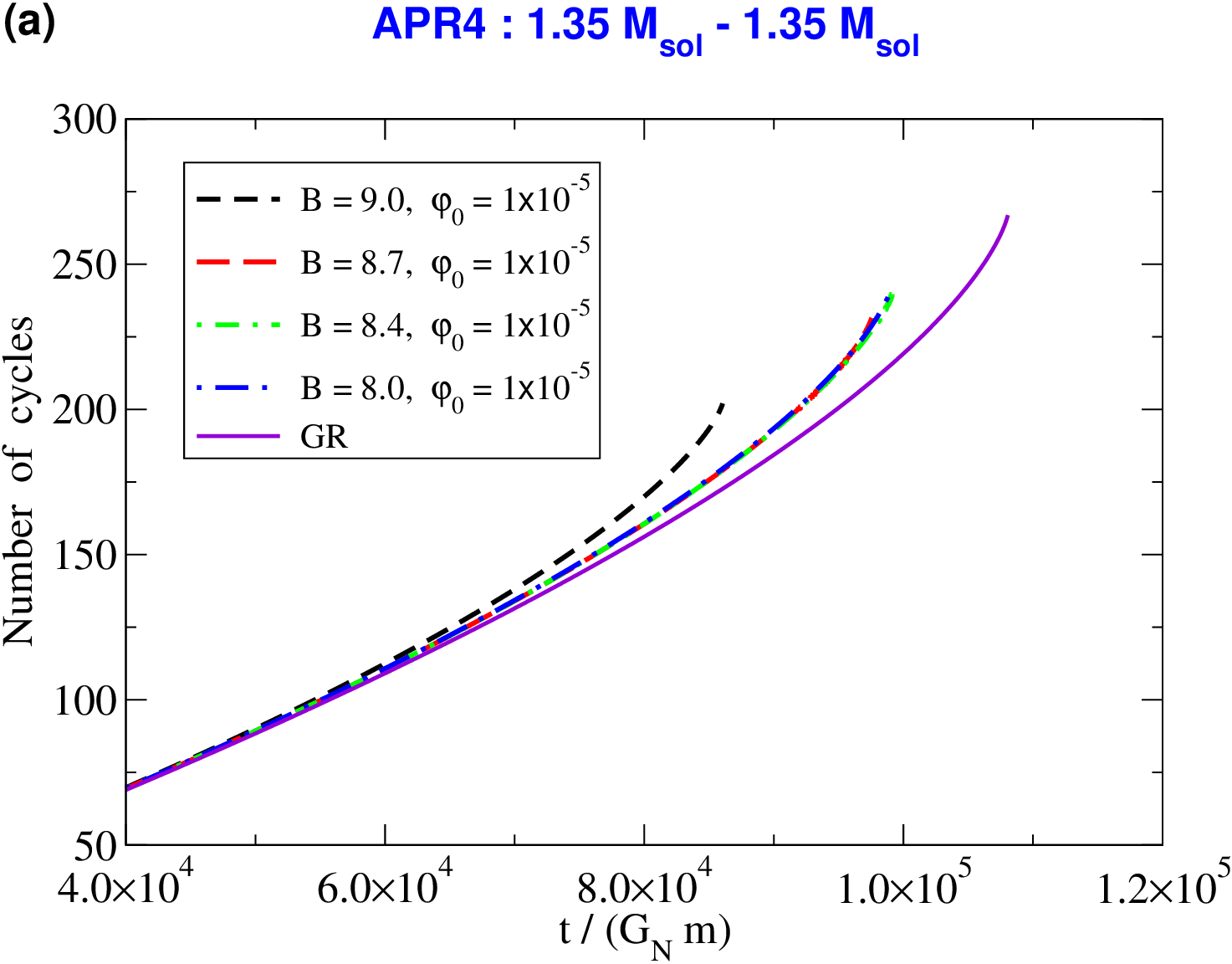} \\
\vspace{0.4cm}
  \includegraphics[width=7.cm]{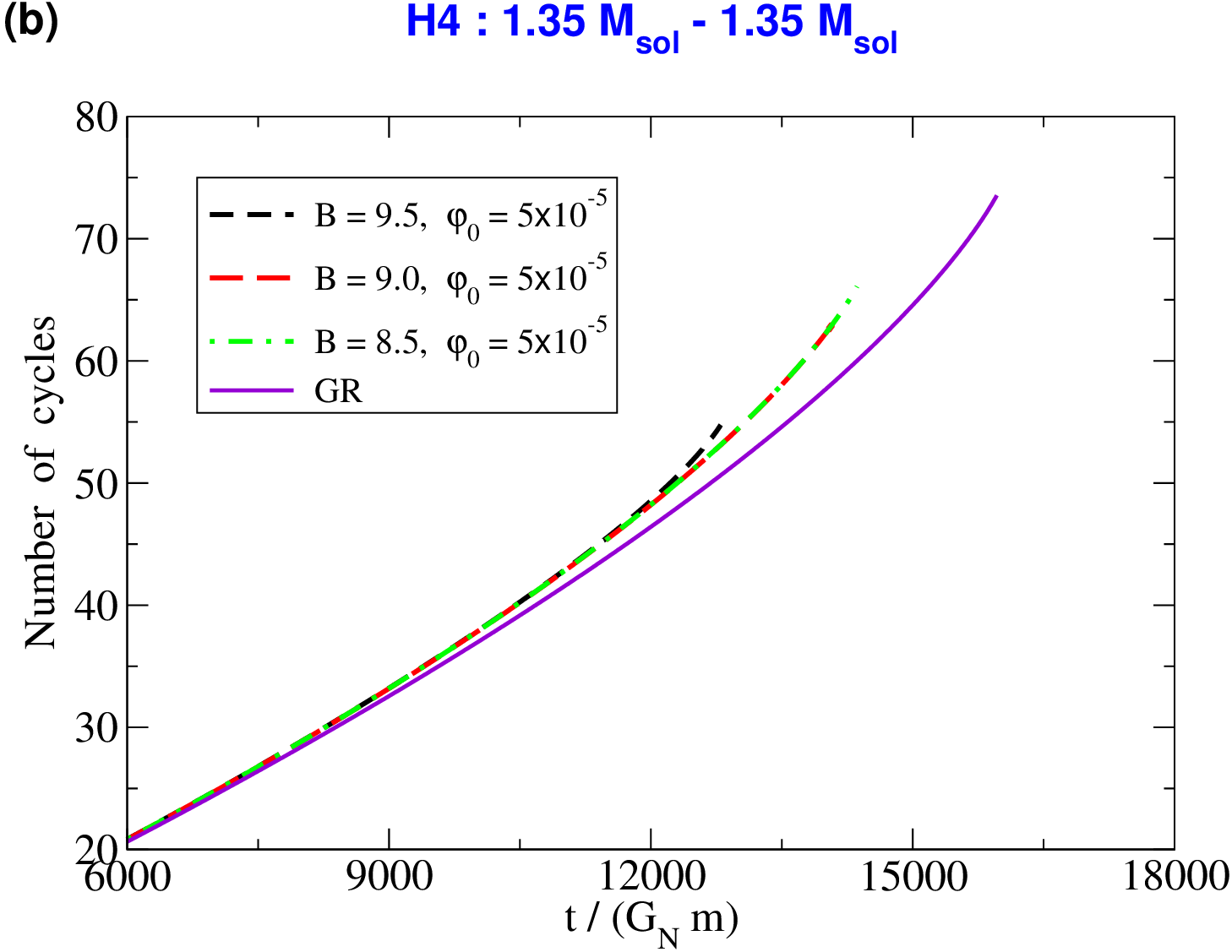}
  \caption{Same as Fig.~\ref{fig8} but for the choice (ii)
    for the gravitational energy flux.
    In the general-relativity case, the 3.5PN GR energy flux is used.
    These are the same data used in Fig.~\ref{fig6} (shown as ``GR'').
    \label{fig9}}
\end{figure}

\begin{table*}
\caption{We list the orbital angular frequencies at the end point of each
quasiequilibrium sequence and the number of GW cycles. We also show
the quantity $\delta {\cal N}_{\rm{GW}}$, which is the difference between
the number of GW cycles in the DEF model and in  general relativity
[i.e., either APR4 (GR) or H4 (GR) depending on the EOS].
When computing the number of cycles, we set the initial orbital
angular frequency to $G_{\rm N} m \Omega =0.005$ ($f_{\rm GW} = 119.7$ Hz)
for the case of APR4 EOS and $G_{\rm N} m \Omega =0.01$ ($f_{\rm GW} = 239.3$ Hz)
for H4 EOS. The subscripts (i) or (ii) refer to the choice of the gravitational
energy flux discussed in Sec. III~D.
\label{table2}}
\begin{ruledtabular}
\begin{tabular}{lccrrrr}
Models & $G_{\rm N} m \Omega_{\rm ener\mbox{-}min}$
& $G_{\rm N} m \Omega_{\rm mass\mbox{-}shed}$
& ${\cal N}_{\rm{GW} (i)}$ & $\delta {\cal N}_{\rm{GW} (i)}$
& ${\cal N}_{\rm{GW} (ii)}$ & $\delta {\cal N}_{\rm{GW} (ii)}$ \\
APR4 (9.0) & 0.0491672 & 0.0565207 & 238.00 & $-28.87$ &   202.03 & $-64.85$ \\
APR4 (8.7) & 0.0494752 & 0.0572480 & 256.62 & $-10.25$ &   231.48 & $-35.40$ \\
APR4 (8.4) & 0.0523407 & 0.0570327 & 262.94 &  $-3.93$ &   239.63 & $-27.24$ \\
APR4 (8.0) & 0.0494658 & 0.0563251 & 265.55 &  $-1.33$ &   242.62 & $-24.25$ \\
APR4 (GR)  & $\cdots$  & 0.0570651 & 266.87 & $\cdots$ & $\cdots$ & $\cdots$ \\
APR4 (3PN) & $\cdots$  & $\cdots$  & 266.08 &  $-0.79$ & $\cdots$ & $\cdots$ \\
APR4 (4PN) & $\cdots$  & $\cdots$  & 265.81 &  $-1.07$ & $\cdots$ & $\cdots$ \\
H4 (9.5)   & 0.0385605 & 0.0438147 & 62.66  & $-10.89$ &   54.76  & $-18.78$ \\
H4 (9.0)   & 0.0401870 & 0.0419199 & 70.89  &  $-2.66$ &   63.68  &  $-9.86$ \\
H4 (8.5)   & $\cdots$  & 0.0410100 & 73.47  &  $-0.07$ &   66.09  &  $-7.46$ \\
H4 (GR)    & $\cdots$  & 0.0405359 & 73.54  & $\cdots$ & $\cdots$ & $\cdots$ \\
H4 (3PN)   & $\cdots$  & $\cdots$  & 74.18  &    0.64  & $\cdots$ & $\cdots$ \\
H4 (4PN)   & $\cdots$  & $\cdots$  & 74.01  &    0.46  & $\cdots$ & $\cdots$ \\
\end{tabular}
\end{ruledtabular}
\end{table*}

As seen in Fig.~\ref{fig3}, because after the onset of dynamical scalarization,
the binding energy calculated in the DEF model decreases less rapidly than
that calculated in general relativity, the binary evolves more quickly
in the DEF model than in general relativity, if differences between
the energy fluxes in general relativity and scalar-tensor theory are neglected.
When an estimate of the scalar-tensor energy flux is included the late
evolution of the binary is even faster (see Figs.~\ref{fig6} and \ref{fig7}).
As a consequence, the number of GW cycles computed from an initial frequency
$f_{\rm GW, ini}$ to a final frequency $f_{\rm GW, fin}$ will be different in 
the DEF model and in general relativity.
To quantify this, we numerically integrate the orbital angular
frequency between the initial and final frequencies discussed in
Table~\ref{table2} and the text around it. When computing the number of
cycles between two models with the same EOS, we impose that 
they have the same initial orbital frequency.
The results are summarized in Table~\ref{table2} and
in Figs.~\ref{fig8} and \ref{fig9}. In particular, the difference in number
of GW cycles between the DEF model
and general relativity for APR4 EOS is 28.9, 10.3, 3.9, and 1.3 for
the case of $B=9.0$, 8.7, 8.4, and 8.0, respectively, if we use the choice (i)
for the gravitational energy flux. On the other hand,
if we adopt the choice (ii), the difference in number of GW cycles
is 64.8, 35.4, 27.2, and 24.3. Note here that the difference is
calculated against the general-relativity case with 3.5PN energy flux
for both choices of (i) and (ii). For the case of H4 EOS, the difference is
10.9, 2.7, and 0.07 for
$B=9.5$, 9.0, and 8.5, respectively, for the choice (i), while for
the choice (ii),
we have 18.8, 9.9, and 7.5, respectively. Thus, except for the case of
H4 EOS $B=8.5$ for 
the choice (i), the difference in number of GW cycles is larger than unity.

However, those numbers should be taken with cautiousness because they are
affected by different source of errors.
For example, the quasiequilibrium configurations themselves
include errors. It is usually common to measure the errors
by a global error indicator, i.e., the error in the virial relation.
In scalar-tensor theory the virial relation is expressed as~\cite{ShibaK13}
\begin{equation}
  M_{\rm Komar} = M_{\rm ADM} + 2 M_{\rm S} \phi_0^{-1},
\end{equation}
where $M_{\rm Komar}$ is the Komar mass. This relation should hold 
along the quasiequilibrium sequences, but because of numerical errors,
deviations can appear. In this paper we define the error in the virial relation
as follows 
\begin{equation}
  {\rm virial~error} = \Bigl| \frac{M_{\rm Komar} - M_{\rm ADM}
    - 2 M_{\rm S} \phi_0^{-1}} {M_{\rm ADM}} \Bigr|.
\end{equation}
In our quasiequilibrium configurations the ``virial error'' is
on the order of $10^{-5}$ for large and medium orbital separations
and $10^{-4}$ for close configurations.
Because the binding energy is on the order of $10^{-3} - 10^{-2}$ 
throughout the computed orbital-frequency range, 
a virial error on the order of $10^{-5} - 10^{-4}$ implies 
that the binding energy has a maximal error of a few \%. 
Besides the error in the quasiequilibrium configurations,
there are errors due to the fitted curves of the binding energy
and the scalar charge. Nevertheless, the difference in the 
number of cycles in Table~\ref{table2} is sufficiently large to make it
worthwhile to run accurate, long full numerical-relativity simulations of binary 
neutron stars in the DEF model and develop accurate template waveforms. 

The frequency region that is affected by dynamical scalarization is
in the several hundreds of Hz, i.e., in the high-frequency portion of
the LIGO/Virgo/KAGRA bandwidth. If the binary is composed by a neutron star
and black hole, dynamical scalarization would in principle take place
at lower frequencies. Quite interestingly, if the binary is in
an eccentric orbit, 
the motion can induce a scalar charge on the black hole~\cite{LiuEWK14}. 
We plan to study in the future whether dynamical scalarization occurs
in a black-hole$-$neutron-star binary and is observable by LIGO/Virgo/KAGRA.

It is worthwhile to note that the difference in number of GW cycles 
between the DEF model and general relativity is much larger
than that between the 3PN approximation of the binding energy and the 4PN one.
(We use the 3.5PN flux for both calculations.) Setting the integration range 
of $G_{\rm N} m \Omega$ to the same as in the general-relativity case,
we obtain that 
the difference in the number of GW cycles between the 3PN approximation and
the 4PN one is 0.28 for APR4 EOS and 0.17 for H4 EOS. (The 4PN case has
a smaller number of cycles than that for 3PN.)

Before closing this section, we would like to comment that the difference
in GW frequency between general relativity and the DEF model was also estimated 
in Ref.~\cite{PalenBPL14}, evolving the 2.5PN equations of motion augmented
by a set of equations that phenomenologically describe the increase of
scalar charge as the two neutron stars come closer to each other
(see Figs.~7, 9 and 11 in Ref.~\cite{PalenBPL14}). An important difference
between the two sets of results is that the quasiequilibrium sequences
in our computation terminate much earlier than those in Ref.~\cite{PalenBPL14}.
This is because the authors of Ref.~\cite{PalenBPL14} treat binary neutron stars
as two spherical neutron stars, while we compute the deformation of the stars
and stop at the mass-shedding point or at the turning point of the binding
energy.

\section{Conclusions}

We have computed quasiequilibrium sequences of binary neutron stars in the
DEF scalar-tensor model~\cite{DamourEF93} that admits dynamical scalarization.
The EOS of the neutron star that we have employed has the form of a piecewise
polytrope and we have used APR4 and H4 EOSs~\cite{ReadLOF09,ReadMSUCF09}. 
We have considered an equal-mass, irrotational binary whose tensor mass at
large separation is $2.7 M_{\odot}$.

Using the quasiequilibrium sequence, we have derived the binding energy 
and scalar charge and found that, as the stars come closer,
and the dynamical scalarization sets in, the binding energy decreases less
rapidly than in general relativity. 
Using the newly computed binding energy and the balance equation,
we have estimated the number of GW cycles during the adiabatic,
quasicircular inspiral stage up to the end of the sequence, which is 
the last stable orbit or the mass-shedding point, depending on which comes
first. When employing the quadrupole component 
of the gravitational energy flux in the scalar-tensor DEF model,
we have found that in the most optimistic case, when dynamical scalarization
sets in around a GW frequency of $\sim 130\, {\rm Hz}$
(i.e., $B = 9.0$ and APR4 EOS), the number of GW cycles from $120$ Hz 
up to merger in general relativity, 
$\sim 270$, is reduced by $24 \%$, of which $11 \%$ is only due to
the inclusion of the scalar-tensor binding energy.
A summary of our results is given in Table~\ref{table2} and
Figs.~\ref{fig6}$-$\ref{fig9} for several choices of the scalar-tensor
parameters. Of course, a reduction in the number of GW cycles 
with respect to the general-relativity case does not immediately inform us 
on whether the deviation can be observed by advanced detectors.
An analysis that take into account the noise spectral density of the
detector and the accumulated 
signal-to-noise ratio would be needed~\cite{Sampetal14}. 
As seen in Table~\ref{table1}, GW frequencies at the onset of
dynamical scalarization are in the several hundreds of Hz, where 
the broadband interferometer configuration of LIGO/Virgo/KAGRA 
has poor sensitivity. In order to measure deviations from 
general relativity in the DEF model, it is crucial that 
the scalar-field parameter $B$ be large so that dynamical scalarization 
sets in at low frequencies (e.g., around $130$ Hz for APR4 EOS and $B=9.0$) 
and large differences in the GW cycles can be observed. 
However, if the parameter $B$ were too large, the DEF model would be 
already rejected by the observational constraints imposed 
by neutron-star$-$white-dwarf binaries~\cite{Freir12,Anton13,BhatBV08}.

Recent studies carried out in Refs.~\cite{Samp13,Sampetal14}, which 
use scalar-tensor templates in the frequency domain, rely on 
the scalar-charge evolution and numerical-relativity simulations of 
Refs.~\cite{BarauPPL13,PalenBPL14}, concluded that advanced detectors
operating at a signal-to-noise ratio (SNR) of 10 will be able to constrain
dynamical scalarization only if the system scalarizes at low enough orbital
frequencies, e.g., $\leq 50$ Hz, so that a sufficient number of GW cycles
emitted during the dynamical-scalarization phase can contribute to the 
accumulated SNR. This would imply that in the case of APR4 EOS with $B=9.0$,
advanced LIGO and Virgo might observe deviations from general relativity
if dynamical scalarization takes place in nature.

Moreover, using results from Ref.~\cite{TanigS10} and from GR computations
shown in this paper (and also a direct integration of the TaylorT4-PN
approximant with tidal effects~\cite{TanjaH}), we find that in general
relativity tidal effects produce a difference of only a few GW cycles
(i.e., $\sim 1 - 3$ GW cycles depending on the EOS) between 130 Hz and
1200 Hz with respect to the point-particle case. Those small differences in
GW cycles induced by tidal effects at high frequency can be measured by
advanced detectors in one single event only if the SNR is roughly
$30 - 35$~\cite{DelPozzo:2013ala,Wade:2014vqa}. Note that depending on the EOS
those differences can be smaller than or comparable to what we have found
in dynamical scalarization (see Table~\ref{table2}). At SNR around $30 - 35$,
deviations from general relativity might also be observable even in cases
in which the onset of dynamical scalarization happens at orbital frequencies
above $50$ Hz~\cite{Samp13,Sampetal14}. It will be interesting to investigate
in the future the detectability of tidal effects in the presence of dynamical
scalarization. To precisely determine for which neutron-star masses, EOS,
and scalar-tensor parameters dynamical scalarization and tidal effects can
be observed with advanced GW detectors, it will be relevant to develop
accurate waveforms in the DEF scalar-tensor model. To this respect the next
work~\cite{SenBST14} is focusing at building accurate analytical templates
that can incorporate dynamical scalarization, and reproduce the binding energy
computed in this paper and the results from numerical-relativity simulations. 

Finally, extending earlier work~\cite{ShibaTOB14}, new long-term numerical
simulations in scalar-tensor theory are suggesting that the analytical
energy flux used in this paper (i.e., the energy flux at quadrupolar order)
is likely to overestimate the exact energy flux in the scalar-tensor DEF model.
Thus, a better modeling of the energy flux (e.g., its PN computation 
through 1PN and even 2PN order~\cite{Lang14}) is crucial for understanding and
quantifying differences from the general-relativity case.

\begin{acknowledgments}
A.B. thanks Noah Sennett for useful discussions on dynamical scalarization.
K.T. acknowledges partial support from JSPS Grant-in-Aid for Scientific
Research (26400267).
M.S. acknowledges partial support from JSPS Grant-in-Aid for Scientific Research
(24244028), JSPS Grant-in-Aid for Scientific Research on Innovative Area
(20105004), and HPCI Strategic Program of Japanese MEXT.
A.B. acknowledges partial support from NSF Grant No. PHY-1208881 and
NASA Grant No. NNX12AN10G.
\end{acknowledgments}


\appendix

\section{Quasiequilibrium sequences of binary systems
with a polytropic equation of state}
\label{app1}

In this Appendix, we work within the DEF scalar-tensor theory and compute
quasiequilibrium sequences for two models of binary neutron stars with
a polytropic equation of state, $P = \kappa \rho^{\Gamma}$, where $\kappa$
is a constant.\footnote{Note that the relation between $\kappa$ and $K$,
which is the polytropic constant defined in Ref.~\cite{BarauPPL13}, is
$\kappa =K c^2$.}
The adiabatic index, $\Gamma$, is set to 2 and the polytropic
constant, $\kappa$, is to $0.0332278$ in units of $c^2/\rho_{\rm nuc}$,
where $\rho_{\rm nuc}$ is the nuclear density defined in
\textsc{LORENE}~\cite{lorene}.
This value of the polytropic constant $\kappa$ is the same as the
one used for the initial data with $\Gamma=2$ that can be
downloaded from the \textsc{LORENE} website~\cite{lorene}.
When we employ physical quantities used in \textsc{LORENE}
(which are in SI units but here we translate them in cgs units),
the nuclear density $\rho_{\rm nuc} = 1.66 \times 10^{14} ~[{\rm g/cm^3}]$,
Newton's constant $G_{\rm N} = 6.6726 \times 10^{-8} ~[{\rm cm^3/(g~s^2)}]$,
the speed of light $c=2.99792458 \times 10^{10} ~[{\rm cm/s}]$, and
the solar mass $M_{\odot} = 1.989 \times 10^{33} ~[{\rm g}]$,
the polytropic constant $\kappa=0.0332278 ~[c^2/\rho_{\rm nuc}]$ is
written as $\kappa/c^2=123.641 G^3 M_{\odot}^2/c^6$. 
This value is slightly larger than the one used by
Barausse {\it et al.}~\cite{BarauPPL13} which is $123G^3 M_{\odot}^2/c^6$. 
(Note that, recently, after our computations finished, 
the units in the \textsc{LORENE} code have been updated. As a consequence 
the fundamental units that we list above differ from the ones 
in the \textsc{LORENE} code by $10^{-4}$.)

We will study the same binary configurations considered in
Ref.~\cite{BarauPPL13}. However, whereas Ref.~\cite{BarauPPL13} computed
the initial data in general relativity, we calculate them in the DEF
scalar-tensor theory. In order to set the values of $B$ and $\varphi_0$,
we need to derive relations between quantities used by
Barausse {\it et al.}~\cite{BarauPPL13} and
Shibata {\it et al.}~\cite{ShibaTOB14}. We will do it using definitions
introduced by Damour and Esposito-Far\`ese~\cite{DamourEF93}. 
We use the subscript {\sc DEF} for Damour and Esposito-Far\'ese,
{\sc BPPL} for Barausse {\it et al.}, and {\sc STOB} for Shibata {\it et al.}.

As given in Ref.~\cite{BarauPPL13}, the relation between
$\varphi_{\rm BPPL}$ and $\varphi_{\rm DEF}$ is
$\varphi_{\rm BPPL} = \varphi_{\rm DEF} / \sqrt{4 \pi G}$, and
the var-type scalar field is related to the scalar field
$\phi$ as $\phi = \exp( -\beta \varphi_{\rm BPPL}^2 )$,
where $\beta$ is a constant.
On the other hand, the relation between
$\varphi_{\rm STOB}$ and $\varphi_{\rm DEF}$ is
$\varphi_{\rm STOB} = \sqrt{B} \varphi_{\rm DEF}$,
and the var-type scalar field is related to the scalar field
as $\phi = \exp( \varphi_{\rm STOB}^2 / 2 )$~\cite{ShibaTOB14}.
From these equations we conclude that the relations
between the definitions by Barausse {\it et al.}~\cite{BarauPPL13} and
Shibata {\it et al.}~\cite{ShibaTOB14} are 
\begin{eqnarray}
  &&B = -2 \Bigl( \frac{\beta}{4 \pi G} \Bigr), \\
  &&\varphi_{\rm STOB} = \sqrt{4 \pi G B} \varphi_{\rm BPPL}.
\end{eqnarray}
Thus, the parameters used by Barausse {\it et al.},
$\beta/ (4 \pi G) = -4.5$ and
$(\varphi_{\rm BPPL})_0 = 10^{-5} G^{-1/2}$,
correspond to
$B = 9.0$ and $(\varphi_{\rm STOB})_0 = \sqrt{4 \pi B} \times 10^{-5}
= 6 \sqrt{\pi} \times 10^{-5}$ 
in this paper. For simplicity, in the following we drop ``STOB''
from $\varphi_{\rm STOB}$.

\begin{figure}
  \vspace{0.5cm}
  \includegraphics[width=7.7cm]{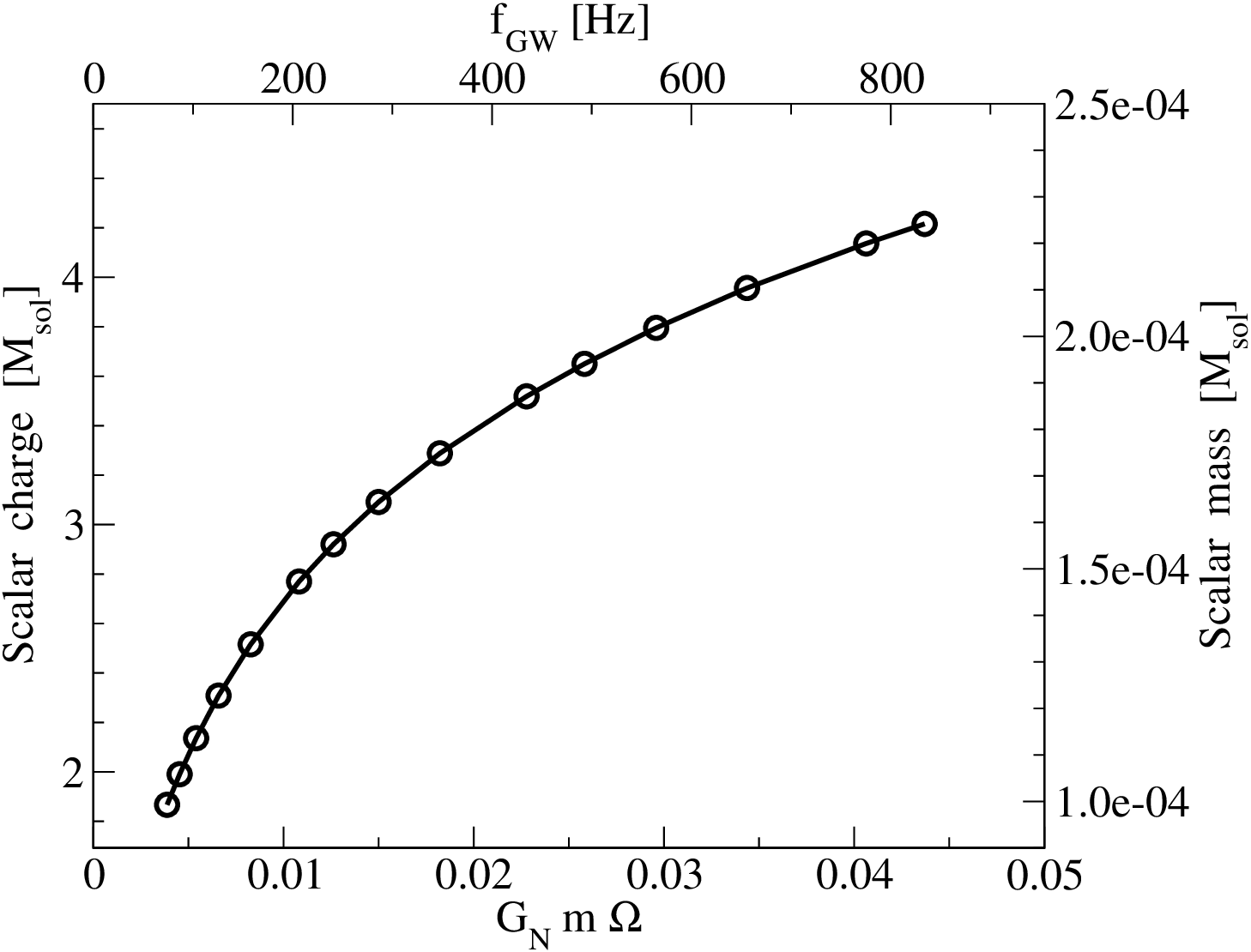}
  \caption{Scalar charge (left $y$ axis) or scalar mass (right $y$ axis)
    as a function of the orbital angular frequency normalized to the
    tensor mass at infinite separation (lower $x$ axis) 
    or as a function of the frequency of GWs 
    defined by $f_{\rm GW} \equiv \Omega/\pi$ from an unequal-mass
    binary neutron star with $m = 3.39 M_\odot$ (upper $x$ axis).
    The equation of state is a polytropic one with $\Gamma=2$.
    Note that the vertical axis is in a linear scale.
    \label{fig10}}
\end{figure}

\begin{figure}
  \vspace{0.5cm}
  \includegraphics[width=8.0cm]{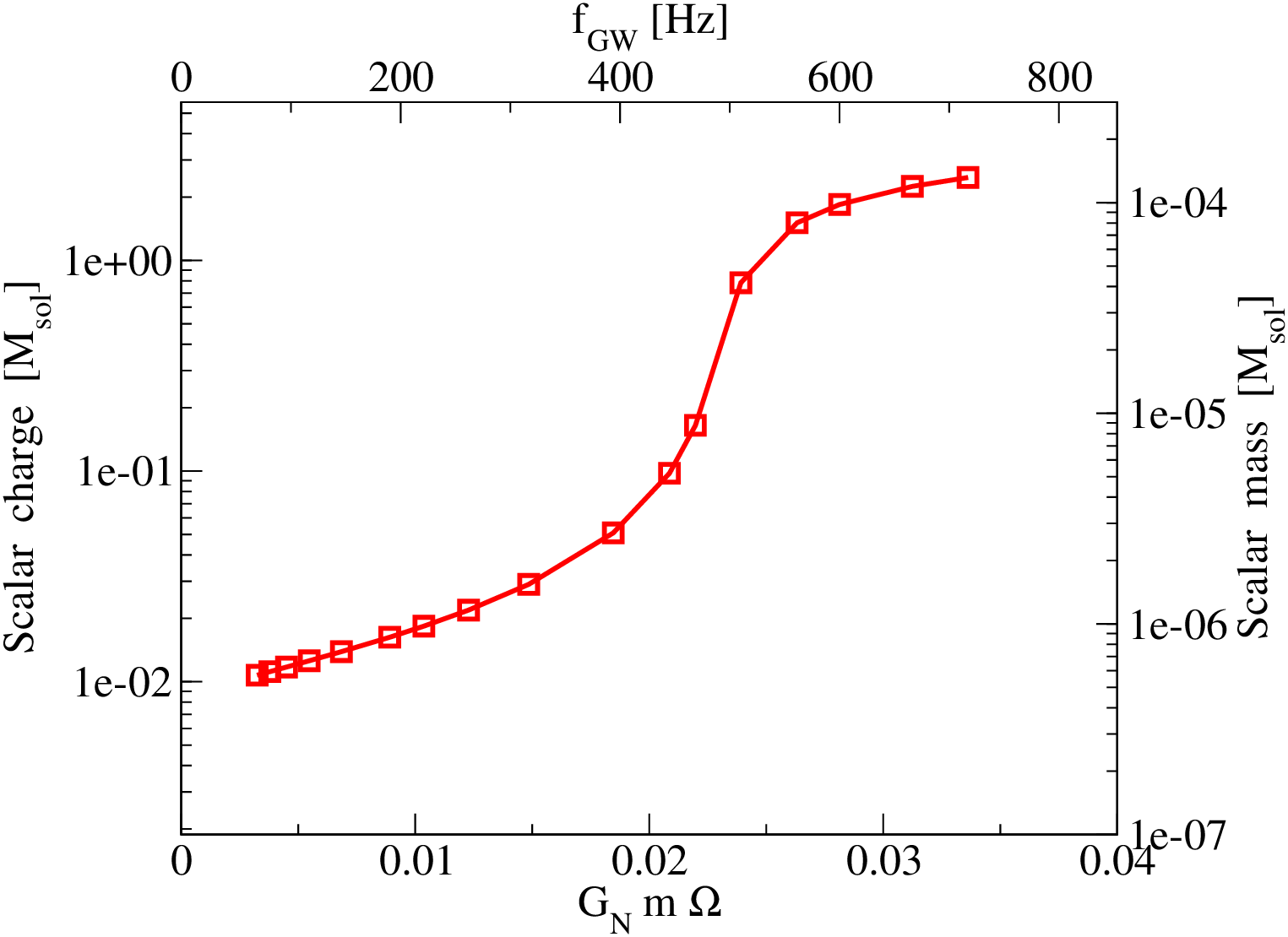}
  \caption{Same as Fig.~\ref{fig10} but for an equal-mass
    binary neutron star with $m = 3.03 M_\odot$.
    Note that the vertical axis is in a logarithmic scale.
    \label{fig11}}
\end{figure}

Fixing the quantities $B$ and $\varphi_0$ to the above values,
we compute two configurations of binary neutron stars. The first one 
describes an unequal-mass binary in 
which the baryonic rest masses of each star are 
$1.78 M_{\odot}$ and $1.90 M_{\odot}$, respectively.
The gravitational masses of spherical stars having the
same baryonic rest masses are $1.64 M_{\odot}$
(with compactness 0.160) and $1.74 M_{\odot}$ (with compactness 0.181).
Thus the tensor mass at infinite separation is $m=3.39 M_{\odot}$.
The more massive star is spontaneously scalarized
when it is a spherical (isolated) star, and it has the scalar charge of
$0.790 M_{\odot}$ and the scalar mass of $4.20 \times 10^{-5} M_{\odot}$.
On the other hand, the less massive star is not spontaneously
scalarized in a spherical (isolated) state and has the scalar charge of
$0.0117 M_{\odot}$ and the scalar mass of $6.22 \times 10^{-7} M_{\odot}$.
Because of the more massive star's scalar field, the binary system is
already scalarized as demonstrated in Fig.~\ref{fig10}
(note the vertical axis in the figure is in linear scale). 
As discussed in Sec.~\ref{intro}, because the binary system is already 
scalarized, the general-relativity initial data used 
in this case by Barausse {\it et al.} have artificially put the binary system 
in a local minimum of the binding energy. Thus, we suspect that the fast plunge 
seen by Barausse {\it et al.} might be enhanced by this effect.
It would be interesting to repeat the simulation using initial data in
the DEF scalar-tensor theory.

The other configuration that we consider is an equal-mass binary in which
the baryonic rest mass of both stars is $1.625 M_{\odot}$.
The gravitational mass of a spherical star with the same
baryonic rest mass is $1.51 M_{\odot}$ and the compactness is 0.140.
The tensor mass at infinite separation is $m=3.03 M_{\odot}$.
The star is not spontaneously scalarized in a spherical (isolated) state. 
Its scalar charge is $4.00 \times 10^{-3} M_{\odot}$ and the scalar mass
is $2.13 \times 10^{-7} M_{\odot}$. So, the binary system is not spontaneously
scalarized at infinite separation. As shown in Fig.~\ref{fig11},
when using the quasiequilibrium sequence, the dynamical scalarization sets 
in at around the GW frequency of
$f_{\rm GW} \sim 450~{\rm Hz}$. Barausse {\it et al.} reported that the
dynamical scalarization occurs at the GW frequency of
$f \sim 645~{\rm Hz}$. Although this value is slightly larger than ours,
those results are consistent. Indeed, we found~\cite{ShibaTOB14} that
the occurrence of dynamical scalarization in a dynamical simulation
tends to be delayed compared with that in a quasiequilibrium sequence.
This might be due to the infall motion in the simulation.

\begin{figure}[!ht]
  \includegraphics[width=7.cm]{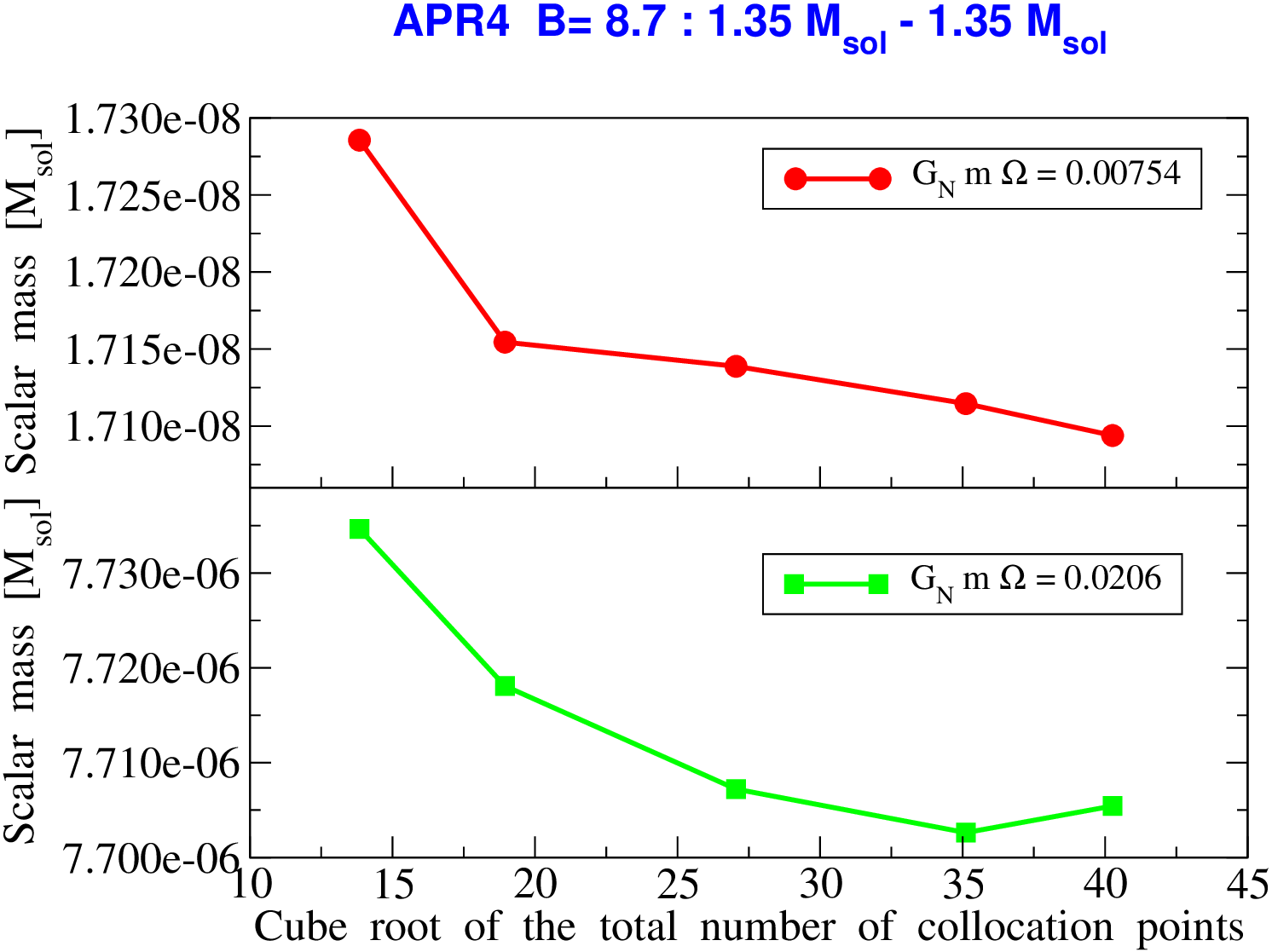} 
  \caption{Scalar mass of APR4 $B=8.7$ as a function
    of the cube root of the total number of collocation points,
    ${}^3 \!\!\!\!\! \sqrt{N_r \times N_{\theta} \times N_{\hat{\varphi}}}$.
    The upper panel is the scalar mass at orbital angular frequency of
    $G_{\rm N} m \Omega =0.00754$, and the lower one is for
    $G_{\rm N} m \Omega =0.0206$.
    \label{fig12}}
\end{figure}

\section{Convergence test for the scalar mass} \label{app2}

In this Appendix we compare the scalar mass of APR4 $B=8.7$
at orbital angular frequency of $G_{\rm N} m \Omega =0.00754$ and
0.0206 for different resolutions (i.e., for different number of
collocation points). Those orbital angular frequencies are before
and after dynamical scalarization. We choose five resolutions,
$N_r \times N_{\theta} \times N_{\hat{\varphi}} =49 \times 37 \times 36$,
$41 \times 33 \times 32$, $33 \times 25 \times 24$,
$25 \times 17 \times 16$, and $17 \times 13 \times 12$,
where $N_r$, $N_{\theta}$, and $N_{\hat{\varphi}}$ are the number of
collocation points for the radial, polar, and azimuthal directions,
respectively.

Figure \ref{fig12} shows the scalar mass of APR4 $B=8.7$ as a function
of the cube root of the total number of collocation points,
${}^3 \!\!\!\!\! \sqrt{N_r \times N_{\theta} \times N_{\hat{\varphi}}}$.
We find that the scalar mass for $33 \times 25 \times 24$
(${}^3 \!\!\!\!\! \sqrt{N_r \times N_{\theta} \times N_{\hat{\varphi}}}
\simeq 27.05$) is in approximately convergent level for the case
of $G_{\rm N} m \Omega =0.0206$ (the case after dynamical scalarization).
The relative deviation from the values for higher resolutions is less
than $1 \times 10^{-3}$. For the case of $G_{\rm N} m \Omega =0.00754$
(the case before dynamical scalarization), the resolution of
$33 \times 25 \times 24$ seems to be not enough because the orbital
separation is twice larger than the case of $G_{\rm N} m \Omega =0.0206$.
However, the relative deviation of the result for
$33 \times 25 \times 24$ from higher resolutions is on the order of
a few $\times 10^{-3}$ in $1.7 \times 10^{-8}$.

From those convergence tests and to save computational time for our
limiting resources, we decide to choose the number of collocation
points of $33 \times 25 \times 24$. For much closer cases just before
the end of quasiequilibrium sequences, we use
$33 \times 21 \times 20$ and $33 \times 17 \times 16$,
keeping the number of collocation points for the radial direction.

\begin{figure*}[!ht]
  \includegraphics[width=7.cm]{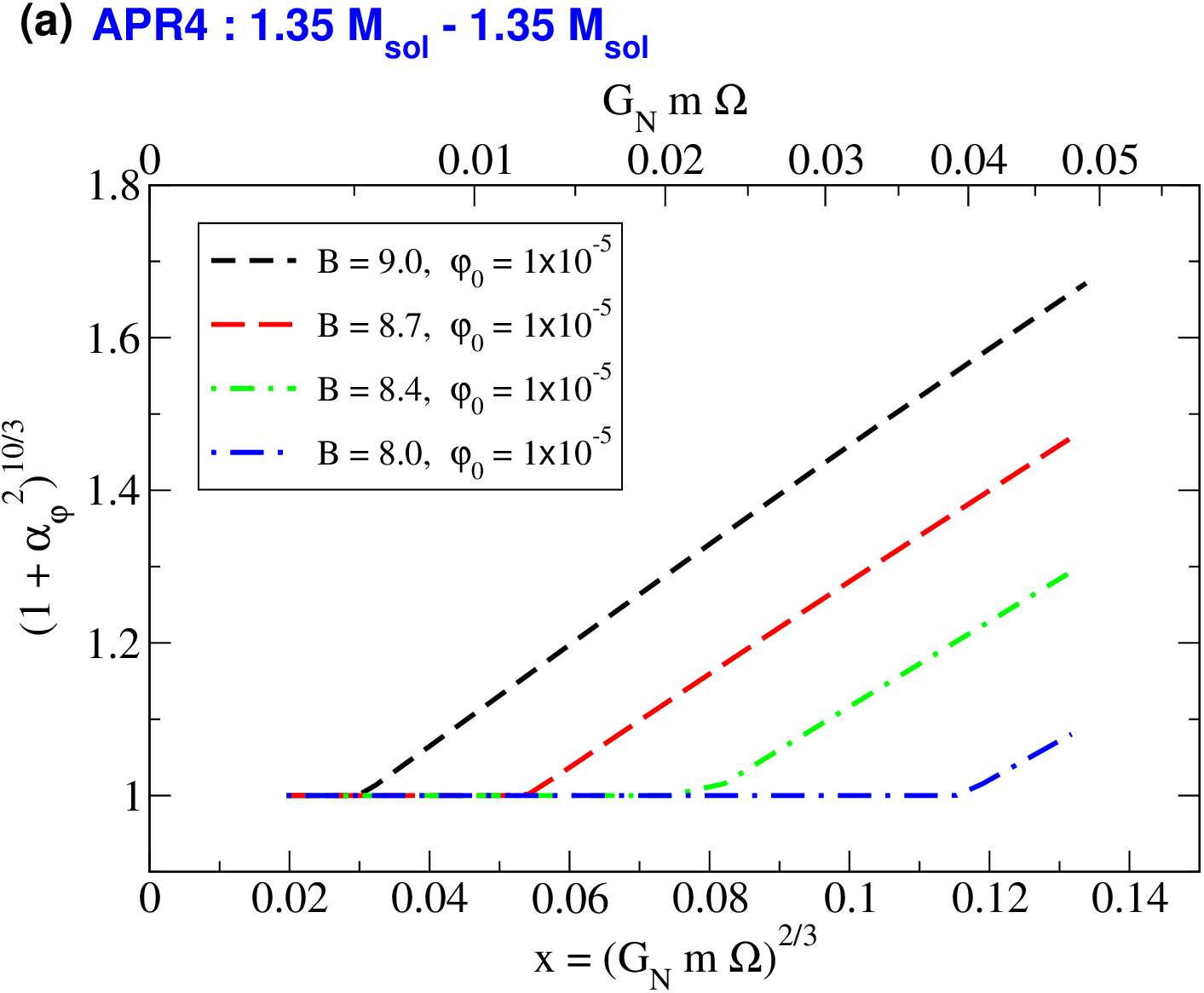} 
\hspace{0.5cm}
  \includegraphics[width=7.cm]{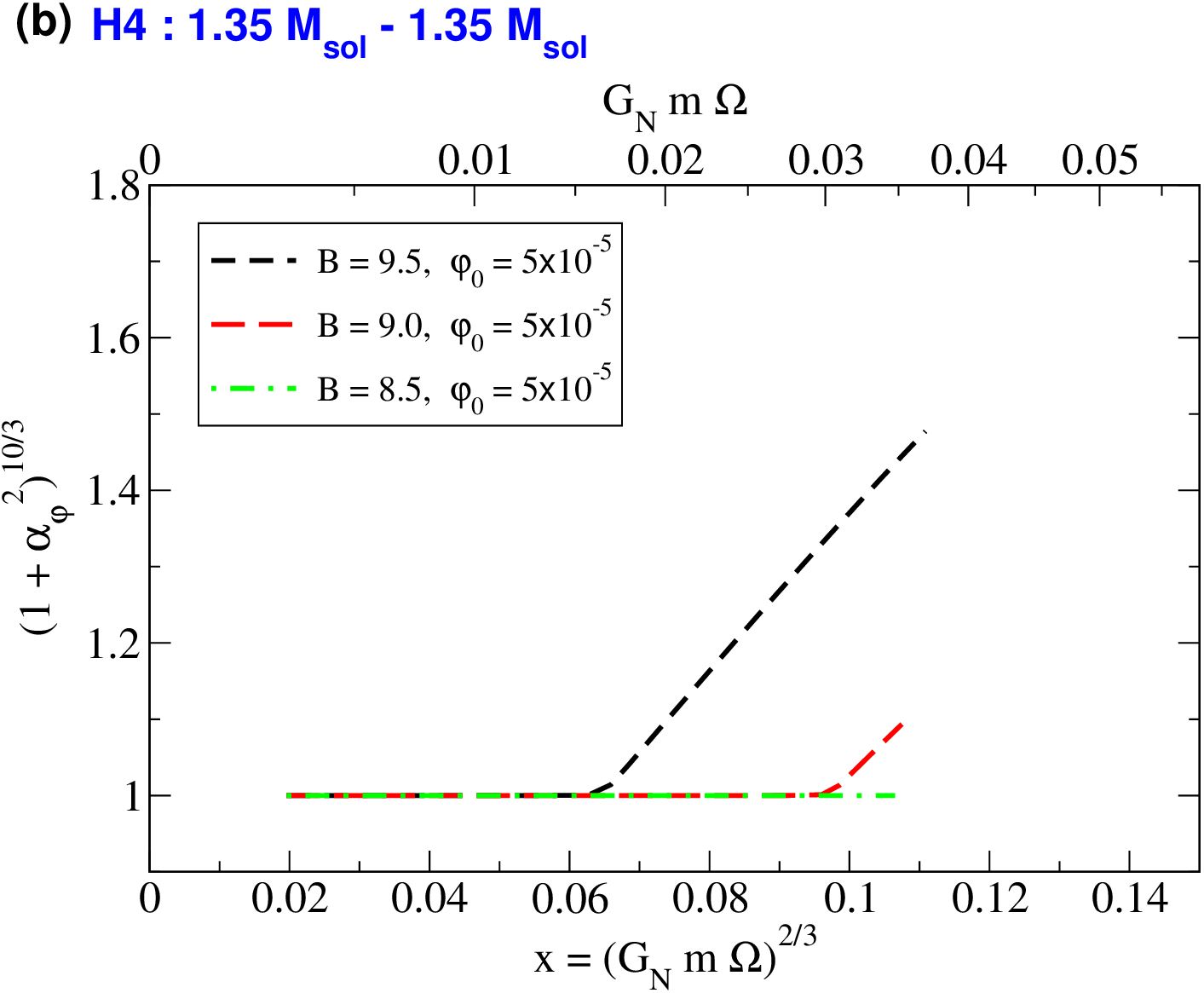}
  \caption{We plot the function $(1 + \alpha_{\varphi}^2 )^{10/3}$ that
    appears in Eq.~(\ref{eq:stflux}) versus 
    $x \equiv (G_{\rm N} m \Omega)^{2/3}$ (lower $x$ axis)
    and the orbital angular frequency normalized
    to the tensor mass at infinite separation (upper $x$ axis).
    Left panel (a) shows results for APR4 EOS. Black short-dashed,
    red long-dashed, green dot-short-dashed, and blue dot-long-dashed curves
    are, respectively, the cases: $B=9.0$, 8.7, 8.4, and 8.0.
    Right panel (b) shows results for H4 EOS. Black short-dashed,
    red long-dashed, and green dot-dashed curves are, respectively,
    the cases: $B=9.5$, 9.0, and 8.5.
    \label{fig13}}
\end{figure*}

\begin{figure*}[!ht]
  \includegraphics[width=7.cm]{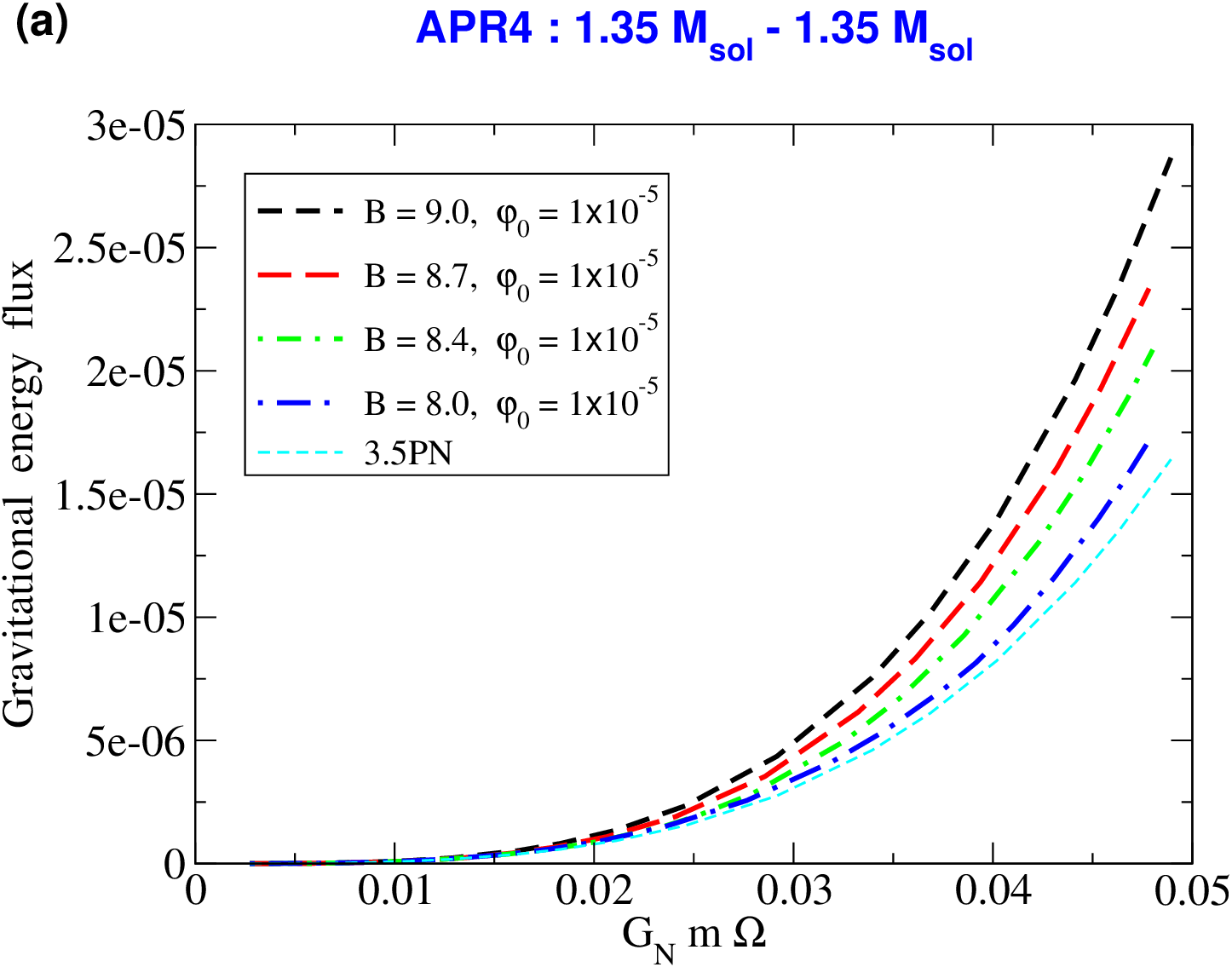} 
\hspace{0.5cm}
  \includegraphics[width=7.cm]{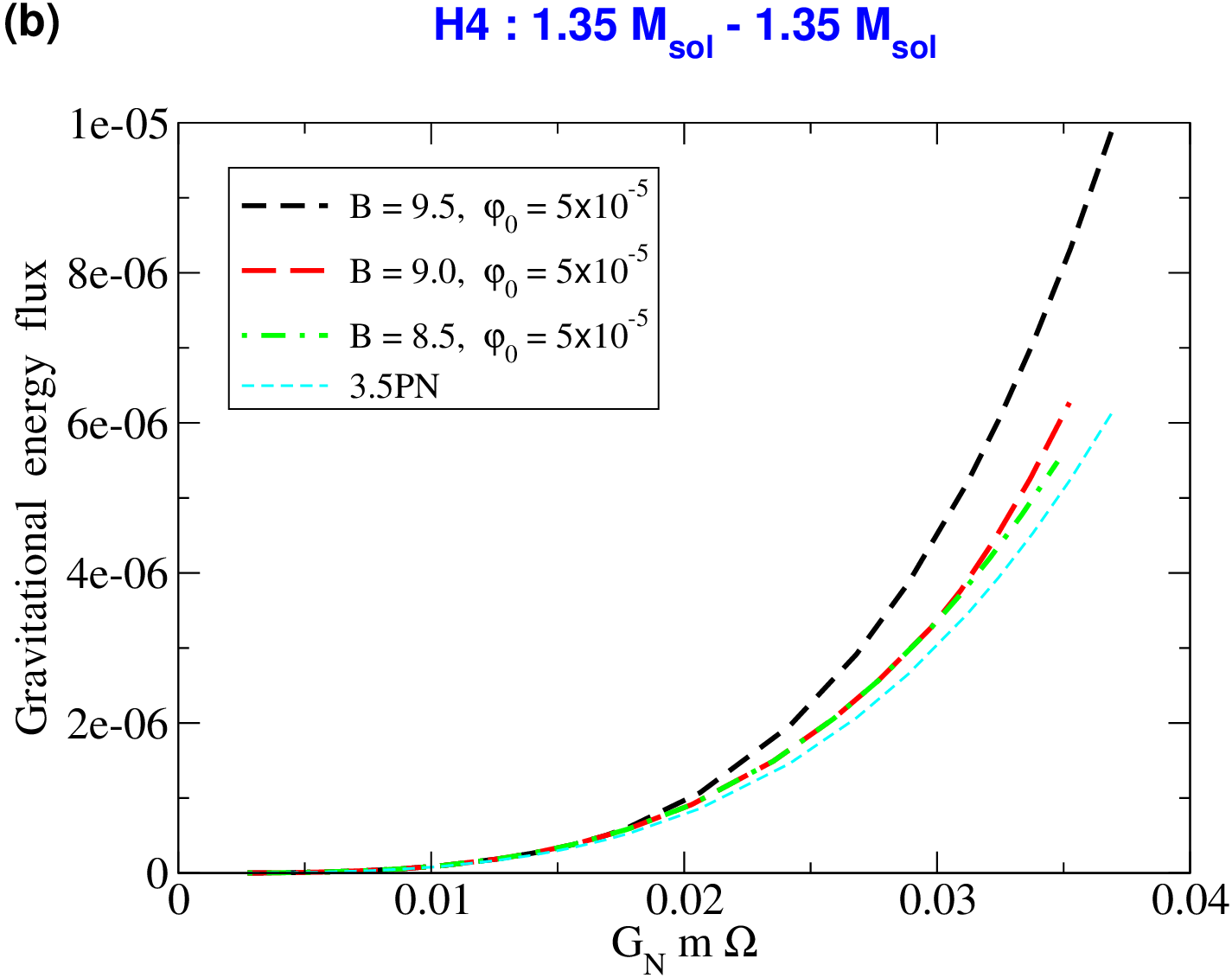}
  \caption{Gravitational-wave energy flux as a function of the
    orbital angular frequency normalized to the tensor mass
    at infinite separation.
    Left panel (a) shows results for APR4 EOS. Black short-dashed,
    red long-dashed, green dot-short-dashed, and blue dot-long-dashed curves
    are, respectively, the cases: $B=9.0$, 8.7, 8.4, and 8.0.
    Right panel (b) shows results for H4 EOS. Black short-dashed,
    red long-dashed, and green dot-dashed curves are, respectively,
    the cases: $B=9.5$, 9.0, and 8.5.
    Those results are calculated by using Eq.~(\ref{eq:stflux}).
    In both panels, the cyan dashed curve refers to the 3.5PN energy flux
    in general relativity.
    \label{fig14}}
\end{figure*}

\section{On the gravitational-wave energy flux} \label{app3}

In this Appendix we first explain how we compute the energy flux used in case 
(ii) (see discussion in Sec.~\ref{sec:evolorb}), 
\begin{equation}
  {\cal F}_{\rm G}^{\rm Quadrupole} = \frac{32}{5} \nu^2 \Bigl[ G_{\rm N} m \Omega
    (1 + \alpha_{\varphi}^2 ) \Bigr]^{10/3}, \label{eq:stflux}
\end{equation}
and then we compare it to the 3.5PN flux in general
relativity~\cite{Blanchet06}.

Quite interestingly, by plotting the function 
$(1 + \alpha_{\varphi}^2 )^{10/3}$ that appears in Eq.~(\ref{eq:stflux}) 
versus $x \equiv (G_{\rm N} m \Omega)^{2/3}$ (see Fig.~\ref{fig13}), 
we find that it can be well approximated by the following simple fit
\begin{equation}
  (1 + \alpha_{\varphi}^2 )^{10/3} = \left\{
  \begin{array}{ll}
    1          & \qquad {\rm before~dyn.~scal.} \\
    a_0 + a_1 x & \qquad {\rm after~dyn.~scal.}
  \end{array}
  \right. \label{eq:scfit}
\end{equation}
where $a_0$ and $a_1$ are constants obtained  
fitting the curves in Fig.~\ref{fig13} after dynamical scalarization.  
As seen in Fig.~\ref{fig1}, the case of H4 EOS $B=8.0$
does not reach dynamical scalarization; thus, in this case we do not
use Eq.~(\ref{eq:scfit}), but adopt a polynomial
fit in $x$. Furthermore, by taking the inverse of Eq.~(\ref{eq:scfit}), 
we can express the scalar charge after dynamical scalarization through 
the analytic formula 
\begin{equation}
  M_{\varphi} = m \sqrt{B} \Bigl[ (a_0 +a_1 x)^{3/10} - 1 \Bigr]^{1/2}.
\end{equation}

Finally, using the above results and Eq.~(\ref{eq:alpha}),
we plot in Fig.~\ref{fig14} ${\cal F}_{\rm G}^{\rm Quadrupole}$ and 
the 3.5PN energy flux in general relativity. As we can see, depending
on the scalar-tensor parameters, the fractional difference between
${\cal F}_{\rm G}^{\rm Quadrupole}$ and the general-relativity flux 
can become $70 \%$. However, as pointed out in Sec.~\ref{sec:scalar},
dynamical scalarization in quasiequilibrium configurations occurs earlier
than in numerical simulations~\cite{ShibaTOB14}. Moreover the increase
in scalar charge computed in the simulations is smaller than that 
obtained in quasiequilibrium study. These facts suggest that the value 
of the energy flux in case (ii) is likely an overestimate of the exact result.
Indeed, as explained in the footnote in Sec.~\ref{sec:evolorb},
the energy flux obtained in long-term simulations seems to lie between 
cases (i) and (ii).

\bibliography{nsnsstRef}

\end{document}